\documentclass[draft, jgrga]{AGUTeX}

\usepackage{lineno}
\linenumbers*[1]

\usepackage{amsmath} 

\usepackage{graphicx}
\graphicspath{{Figures/}}		
\setkeys{Gin}{draft=false}
\authorrunninghead{PILKINGTON ET AL.}

\titlerunninghead{POLAR CONFINEMENT}

\authoraddr{Corresponding author: N. M. Pilkington,
Department of Physics and Astronomy, University College London, Gower St., London, WC1E 6BT, UK.
(nathanp@star.ucl.ac.uk)}

\begin{document}

%
%

\title{Polar Confinement of Saturn's Magnetosphere Revealed by in-situ Cassini Observations}
%
%

%
%



\authors{N. M. Pilkington,\altaffilmark{1,2}
N. Achilleos,\altaffilmark{1,2,3} C. S. Arridge,\altaffilmark{4,2}
A. Masters,\altaffilmark{3} N. Sergis,\altaffilmark{5} A. J. Coates,\altaffilmark{4,2} and M. K. Dougherty\altaffilmark{6}}

\altaffiltext{1}{Department of Physics and Astronomy, University College London, Gower St., London, WC1E 6BT, UK.}

\altaffiltext{2}{Atmospheric Physics Laboratory, The Centre for Planetary Sciences at UCL/Birkbeck, Gower St., London, WC1E 6BT, UK.}

\altaffiltext{3}{Institute of Space and Astronautical Science, Japan Aerospace Exploration Agency, Sagamihara, Kanagawa, Japan.}

\altaffiltext{4}{Mullard Space Science Laboratory, Department of Space and Climate Physics, University College London, Dorking, UK.}

\altaffiltext{5}{Academy of Athens, Office of Space Research \& Technology, Athens, Greece.}

\altaffiltext{6}{Blackett Laboratory, Imperial College London, London, UK.}

%
%


\begin{abstract}
Plasma rotation plays a large role in determining the size and shape of Saturn's disc-like magnetosphere. A magnetosphere more confined to the equator in the polar regions is expected as a result of the interaction between this type of obstacle and the solar wind. In addition, at times away from equinox, a north-south asymmetry is expected where the magnetopause will be further confined in one hemisphere but less confined in the opposite hemisphere. Examining the extent of this confinement has been limited by a lack of high-latitude spacecraft observations. Here, for the first time, direct evidence for polar confinement of Saturn's magnetopause has been observed using in-situ data obtained by the \textit{Cassini} spacecraft during a series of high-inclination orbits between 2007 and 2009. Following techniques established by previous authors, we assume an equilibrium between the solar wind dynamic pressure (which \textit{Cassini} is generally unable to measure directly), and the magnetic plus plasma pressure inside the magnetosphere. This assumption thus allows us to estimate the upstream solar wind dynamic pressure ($D_\textnormal{P}$) for a series of magnetopause crossings, and hence to determine the expected location and global shape of the magnetopause as a function of $D_\textnormal{P}$.\\
A clear divergence from the familiar axisymmetric models of the magnetosphere is observed, which may be characterised by an `apparent flattening parameter' of 0.81+0.03/-0.06 (representing a simple dilation of the nominal axisymmetric boundary along the $Z_{\textnormal{KSM}}$ axis such that the extent is reduced by approximately 19\% in this direction). This figure is insensitive to variations in $D_\textnormal{P}$.
\end{abstract}

%
%

%

\begin{article}

%
%

\section{Introduction}
The solar wind is a supersonic plasma which continuously flows away from the Sun and fills the entire solar system. Its formation was theorised by \citet{Parker1958} and it was first directly observed by the Luna 1 spacecraft and confirmed by \citet{Snyder1963} using Mariner 2 observations. It carries with it a remnant of the solar magnetic field, known as the Interplanetary Magnetic Field (IMF). When it encounters a magnetised body it is slowed, heated and deflected around that obstacle. The resulting cavity, to which the solar wind cannot obtain direct access, is known as the magnetosphere.\\
The magnetopause is the current sheet boundary that separates the plasma populations and magnetic fields of solar and planetary origin and defines the area within which forces internal to the magnetosphere dominate the ram pressure of the solar wind. The solar wind ram, or dynamic, pressure is highly variable and has a strong influence on the magnetosphere, whose size and shape can exhibit rapid variability.\\
Magnetospheres vary greatly in their global structure depending chiefly on the strength of the planetary magnetic field, the amount of internal plasma and the distance from the Sun. The magnetosphere of Mercury, for example, is relatively small and can barely hold off the solar wind from the planet's surface \citep[e.g.,][]{Slavin2007}, whereas the magnetosphere of Jupiter is the largest structure in the solar system \citep[e.g.,][]{Bagenal1992}. This is due, in part, to its strong magnetic field and also due to the presence of a highly volcanically active moon, Io, which was found to release approximately 1000\,kg\,$s^{-1}$ of sulphur dioxide gas into the magnetosphere \citep[e.g.,][]{Dessler1980}. This is then partly ionised, through charge exchange, into a plasma and is picked up by the rotating ambient plasma flow in the vicinity of the moon and acts to significantly inflate the magnetosphere due to enhanced plasma pressure. The solar wind dynamic pressure is also much smaller at Jupiter's orbital distance. \\
In terms of size, the Earth's magnetosphere lies somewhere between the extreme examples previously discussed. It doesn't have a large, internal source of plasma as Jupiter does, but it experiences a much smaller dynamic pressure than Mercury's magnetosphere owing to its larger distance from the Sun. A statistical study by \citet{Achilleos2008} found that Saturn's magnetospheric size follows a bimodal distribution (the sum of two Gaussian distributions) with the most common stand-off distances at $\sim$22 and $\sim$27 Saturn planetary radii ($R_\textnormal{S}$). The Kronian magnetosphere thus falls somewhere between Earth and Jupiter in terms of size and exhibits a similar 'dual state' to the Jovian system \citep{Joy2002}.  \\
Plasma loading plays a significant role in shaping the Kronian magnetosphere and Enceladus is the dominant source of plasma in this system. It is thought to be cryovolcanic, its volcanism arising from the tidal heating of its interior by Saturn. \citet{Tokar2006} and \citet{Pontius2006} estimate that 100\,kg\,$s^{-1}$ of water molecules are liberated by the moon, \citet{WaiteJr2009} found that smaller quantities of carbon dioxide, ammonia and hydrocarbons are also present in the Enceladus plasma. More recent estimates made by \citet{Spencer2011} place the mass outflow rate to be as high as 200\,kg\,$s^{-1}$. \\
The water molecules are then partially ionised and the presence of this outflowing plasma acts to inflate the magnetosphere significantly. The rapid rotation of the planet confines it to the equatorial plane and forms a magnetodisc. It then diffuses into the outer magnetosphere due to centrifugal instabilities as described by \citet{Kivelson2005}. Only the ions and electrons with sufficient energy can escape and travel along the field lines to higher latitudes. \\
The obstacle presented to the solar wind flow is thus disc-like in nature and is more streamlined than the relatively blunt obstacle presented by the terrestrial magnetosphere. The solar wind flows more easily over the polar regions of a disc-like magnetosphere and thus some degree of polar flattening of the magnetopause is expected. \\
At times when the dipole moment of the planet is tilted with respect to the solar wind flow (i.e. no longer perpendicular), a north-south asymmetry (or 'hinging') in the magnetosphere is formed. As a result, the magnetosphere may appear further confined in one hemisphere whereas an apparent inflation may be observed in the other hemisphere. The angular separation between the magnetic dipole and the solar wind direction varies seasonally so the magnitude of this effect is also thought to vary with planetary season. It is difficult to distinguish between these sources of confinement, and studies at different planetary season may be the only way to unambiguously separate them. \\
An additional effect is that of the magnospheric oscillation observed at Saturn by e.g. \citet{Espinosa2000, Espinosa2001, Cowley2006, Kurth2008, Andrews2008, Clarke2010, Arridge2011, Provan2011, Andrews2012, Provan2012, Provan2013}. This oscillation seems to be strongly linked to the phase of the Saturn Kilometric Radiation (SKR) and appears to be caused by a current system which rotates with the planet. It has been suggested by \citet{Espinosa2003} that a compressional wave is generated close to the planet by an equatorial magnetic anomaly which causes a periodic change in the magnetic field with a period close to that of the planet's rotation. \citet{Clarke2006a} observed planetary-period oscillations in the magnetopause boundary of amplitude 1-2\,$R_\textnormal{S}$. As such, it is important to consider the effects of this anomaly on the structure of the magnetopause and its impact on its high-latitude structure. \\
Polar flattening of the Jovian magnetosphere was observed by \citet{Huddleston1998} using data from the \textit{Galileo}, \textit{Ulysses}, \textit{Voyager}s 1 and 2 and \textit{Pioneer}s 10 and 11. They concluded that the disk-like shape of the obstacle was the cause of the apparent flattening along the north-south axis. Similar results were found by \citet{Joy2002}, who used a combination of observations made by the same spacecraft and MHD simulations, and found that the degree of asymmetry varied with solar wind dynamic pressure. \\
\citet{Huddleston1998} found a power-law relation between the magnetopause stand-off distance and the upstream dynamic pressure just as \citet{Shue1997} found one for the terrestrial magnetosphere,
\begin {linenomath}
\begin{equation}
\label{eq:power_law}
r_0 \propto D_\textnormal{P}^{-\frac{1}{\alpha}}
\end{equation}
\end {linenomath}
where $r_0$ is the magnetopause stand-off distance, $D_\textnormal{P}$ is the solar wind dynamic pressure and $\alpha$ = 6 for a dipole-like configuration as was found for the Earth by \citet{Shue1997}. \citet{Huddleston1998} found a value of $\alpha$ between 4 and 5 for Jupiter, indicating that the Jovian magnetosphere is more compressible and the magnetopause stand-off distance reacts more strongly to changes in the dynamic pressure than the terrestrial magnetosphere. \\
A similar study for the Kronian magnetosphere was made by \citet{Arridge2006} and is the basis of the current study. They used boundary crossings from the first six orbits of \textit{Cassini} to build a shape model of the equatorial magnetosphere and investigate how the size of the magnetosphere reacts to changes in the solar wind dynamic pressure. They found that the size of the magnetosphere could also be described by a power law. A characteristic value of $\alpha$ close to that previously found for Jupiter of {4.3 $\pm$ 0.4} was found. A similar study made by \citet{Achilleos2008} found a value of {5.17 $\pm$ 0.30}. \\
\citet{Kanani2010} built upon this by using a model that included the plasma pressure contributions of electrons and suprathermal ions and also used a more realistic expression for the thermal solar wind pressure. They found a value of $\alpha$ of {5.0 $\pm$ 0.8}. This implies an intermediate compressibility between that of the Jovian and the terrestrial magnetosphere, and this has also been confirmed by \citet{Jia2012} using a global MHD simulation of the Kronian magnetosphere. \\
This study extends this previous work to the high-latitude magnetosphere. Previous studies of Saturn's magnetopause have involved near-equatorial spacecraft orbits only and, hence, have been unable to make any direct assessment of the extent of its polar flattening. As such, the work presented herein represents the most complete picture of Saturn's magnetosphere to date. \\
In Section \ref{sec:model} we discuss the model used to undertake this study and in Section \ref{sec:data} the data used to find the boundary crossings to which the new model is fitted. The results of fitting the new model to the data are reported in Section \ref{sec:results} and in Section \ref{sec:skr} we consider the phase of Saturn's global magnetic oscillation at each point that the spacecraft crosses the magnetosphere in order to determine whether this could be the cause of the observed confinement. Finally, the results of this study are discussed and conclusions are drawn in Section \ref{sec:discussion}.

\section{The model}
\label{sec:model}
\subsection{Previous work}
This study builds on the work of \citet{Arridge2006} and \citet{Kanani2010} who modelled the magnetopause of Saturn by assuming pressure balance between the solar wind dynamic pressure and the magnetic pressure at the magnetopause boundary.  In reality, the magnetopause is unlikely to ever be in true equilibrium, but when considering the average behaviour of the magnetopause over many spacecraft orbits, this is a reasonable assumption to make. \\
In these investigations and the present study, the location of the magnetopause boundary was found using in-situ data from the fluxgate magnetometer onboard the \textit{Cassini} spacecraft, as documented by \citet{Dougherty2002}. \citet{Arridge2006} identified magnetopause crossings within the first six orbits of \textit{Cassini}, between 28 June 2004 and 28 March 2005. However, the high-latitude structure of the Kronian magnetosphere could not be investigated due to the limited coverage of these equatorial orbits. \\
In the absence of an upstream pressure monitor at Saturn, the solar wind dynamic pressure may be estimated by measuring the magnetic field inside the magnetopause and assuming that the total solar wind pressure, $P_{\textnormal{SW}}$, balances the corresponding magnetic pressure,
\begin {linenomath}
\begin{equation}
\label{eq:mag_press}
P_{\textnormal{SW}} = \frac{B^{2}}{2\mu_{0}}\cos^2\Psi
\end{equation}
\end {linenomath}
where $B$ is the magnetic field just inside the magnetopause, ${\mu_{0}}$ is the magnetic permeability of free space and $\Psi$ is the angle between the anti-solar wind direction and the normal to the magnetopause. \\
$P_{\textnormal{SW}}$ consists of two individual (but related) pressure contributions, the dynamic and static pressures,
\begin {linenomath}
\begin{equation}
D_\textnormal{P} + P_0 = \frac{B^{2}}{2\mu_{0}}\cos^2\Psi
\end{equation}
\end {linenomath}
where $P_0$ is the static pressure and $D_\textnormal{P}$ denotes the dynamic pressure, the relative importance of these contributions depends on which part of the magnetopause is being considered. \citet{Arridge2006} set $P_0$ to a constant pressure of $10^{-4}$\,nPa found from average solar wind values, using the work of \citet{Slavin1985a}. However, the solar wind flows in curved streamlines around the magnetosphere which acts to reduce the pressure as opposed to the situation where the particles impact the boundary directly. \citet{Petrinec1997} showed that applying Bernoulli's equation along the solar wind streamlines yields,
\begin {linenomath}
\begin{equation}
\label{eq:Arridge}
\frac{B^{2}}{2\mu_{0}} = kD_\textnormal{P}\cos^2\Psi + P_{0}\sin^2\Psi
\end{equation}
\end {linenomath}
$k$ is the ratio of the pressure at the sub solar point to the upstream solar wind pressure \citep{Kanani2010} and is a factor that relates to how much the dynamic pressure is reduced when the plasma is considered to be flowing along streamlines. As the solar wind flows into the bow shock boundary and around the magnetopause, the streamlines diverge and the plasma is spread out as it flows around the obstacle and hence the flux of momentum across a given area is reduced. A value of 0.881 is appropriate in the case of a supersonic plasma.\\
The dynamic pressure dominates at small values of $\Psi$ which corresponds to the nose of the magnetosphere. The static pressure dominates along the flanks of the magnetosphere where $\Psi \rightarrow 90^\circ$. \\
The formula of \citet{Shue1997} is used to model the magnetopause,
\begin {linenomath}
\begin{equation}
r = r_{0}(D_\textnormal{P}, B_\textnormal{Z}) \left( \frac{2}{1 + \cos\theta} \right) ^{K(D_\textnormal{P}, B_\textnormal{Z})}
\end{equation}
\end {linenomath}
where $r$ is the distance from the centre of the planet to a point on the magnetopause surface, $\theta$ is the angle between the point and the planet-Sun line, $r_0$ is the magnetopause stand-off distance and $K$ is an exponent that controls the flaring of the magnetopause. This formalism is versatile as it can represent a variety of different magnetosphere morphologies. $K < 0.5$ represents a closed magnetosphere and $K > 0.5$ represents an open magnetosphere.\\
The original formalism developed by \citet{Shue1997} to model the terrestrial magnetosphere included dependencies on the $B_\textnormal{Z}$ component of the IMF as well as the solar wind dynamic pressure. Since the IMF isn't thought to play a significant role in determining the size and shape of the magnetospheres of the outer planets, the relations were adapted by \citet{Arridge2006} into the form below,
\begin {linenomath}
\begin{equation}
\label{eq:R0}
r_0 = a_{1}D_\textnormal{P}^{-a_2}
\end{equation}
\begin{equation}
K = a_3 + a_4D_\textnormal{P}
\end{equation}
\end {linenomath}
the coefficients $a_i$ were found by using a non-linear least squares fitting method to fit the the model magnetopause surface to the positions of the observed crossings. This procedure is performed iteratively, the coefficients found in the current iteration are used as a starting point in the next. The coefficients change with each successive iteration until they converge to within a tolerance of $10^{-6}$. \\
The final estimates for the coefficients can in some cases depend on the initial estimates supplied to the solver. This issue is easily solved by repeating the fitting with different starting values. \citet{Arridge2006} found the coefficients to be very stable and within their estimated uncertainties. \\ 
\citet{Kanani2010} improved on this model in several key ways. Firstly, the fixed static pressure was replaced by a more realistic form dependent on dynamic pressure. The ideal gas law is used to do this,
\begin {linenomath}
\begin{equation}
P_0 = nk_\textnormal{B}T
\end{equation}
\end {linenomath}
where $n$ and $T$ are the number density and temperature of the solar wind and $k_\textnormal{B}$ is the Boltzmann constant. An expression for the dynamic pressure is then substituted,
\begin {linenomath}
\begin{equation}
D_\textnormal{P} = \rho u_{\textnormal{SW}}^2
\end{equation} 
\end {linenomath}
where $u_{\textnormal{SW}}$ is the upstream solar wind velocity. \\
The effects of plasma pressure are also included in the model of \citet{Kanani2010}. The \textit{Cassini} electron plasma spectrometer (CAPS-ELS), as documented by \citet{Young2004}, was used to find the pressure associated with electrons of energies between 0.8\,eV and 27\,keV. Corresponding suprathermal ion pressures were found using data from \textit{Cassini}'s Magnetospheric Imaging Instrument (MIMI) as documented by \citet{Krimigis2004}, which is capable of detecting ions with energies in the range 27-4000\,keV. \\
Making these modifications, Equation \ref{eq:Arridge} becomes,
\begin {linenomath}
\begin{equation}
\label{eq:Kanani}
kD_\textnormal{P}\cos^2(\Psi) + \frac{k_\textnormal{B}T_{\textnormal{SW}}}{1.16m_pu_{\textnormal{SW}}^2}D_\textnormal{P}\sin^2(\Psi) = \frac{B^{2}}{2\mu_{0}} + P_{\textnormal{MIMI}} + P_{\textnormal{ELS}}
\end{equation}
\end {linenomath}
where $P_{\textnormal{MIMI}}$ is the pressure contribution of the suprathermal ions measured by the MIMI instrument and $P_{\textnormal{ELS}}$ is the pressure contribution of the electrons measured by the CAPS-ELS instrument. The factor of 1.16 has been introduced to account for the 4\% abundance of $\textnormal{He}^{2+}$ in the solar wind which has a temperature approximately four times greater than the protons, as found by \citet{Slavin1985a}. \\
\citet{Kanani2010} explored the sensitivity of the model to the value of this factor, as well as the solar wind speed and the solar wind temperature but found that it was insensitive to varying these parameters within reasonable limits.

\subsection{Present Study}
In this study, the polar confinement of Saturn's magnetosphere is quantified using a set of magnetopause crossings between early 2007 and late 2008 in order to produce a more complete picture of the dayside magnetopause of Saturn, including, for the first time, its high-latitude structure. We use the techniques employed by \citet{Kanani2010} in order to estimate the solar wind dynamic pressure in the absence of a dedicated upstream solar wind monitor. \\
Comparisons between the dynamic pressure calculated assuming pressure balance and that found by fitting the model through the exact location of each magnetopause crossing are used in order to down-select the data and ensure that the magnetopause is close to equilibrium at the time the spacecraft crosses the magnetosphere. \\
Finally, the phase of the global magnetic oscillation at Saturn is determined for each of the magnetopause crossings in order to provide a preliminary check to determine if the apparent flattening observed is predominantly a result of the global magnetic oscillations (see Section \ref{sec:skr} and the references therein) know to occur throughout the magnetosphere of Saturn. \\
In addition, we have considered the pressure contribution associated with the centrifugal force at the magnetopause as follows. A unit cross-section of the magnetopause layer has centripetal force,
\begin {linenomath}
\begin{equation}
F_{\textnormal{CP}} = mVR\omega^2
\end{equation}
\end {linenomath}
where $m$ is the mass per unit volume, $V$, $\omega$ is the angular velocity of the layer and $R$ is the planet-layer distance. Considering the unit volume of the layer to be the unit area multiplied by the width of the layer then, by definition,
\begin {linenomath}
\begin{equation}
F_{\textnormal{CP}} = \Delta F_{\textnormal{CF}}
\end{equation}
\end {linenomath}
where $\Delta$ is the width of the layer and $F_{\textnormal{CF}}$ is the centrifugal force of the layer. Assuming, generously, that the magnetopause layer has similar density and rotation rate to the plasma just inside the magnetopause, the net force acting on this layer per unit area must supply the centripetal force in order to keep the plasma within the layer rotating, hence,
\begin {linenomath}
\begin{equation}
\frac{B^2}{2\mu_0} + P_{\textnormal{MIMI}} - P_{SW} = -F_{\textnormal{CF}}\Delta
\end{equation}
\end {linenomath}
From \citet{Achilleos2010a} Figure 10 (lower panel), the centrifugal force per unit volume just inside the magnetopause is,
\begin {linenomath}
\begin{equation}
F_{\textnormal{CF}} \sim 3\cdot10^{-9}\left(\frac{{B_0}^2}{\mu_0R_S}\right)
\end{equation}
\end {linenomath}
where $B_0$ is the equatorial surface magnetic field strength with a typical value of \hbox{$\sim$20000\,nT} and $R_\textnormal{S}$ is the equatorial radius of Saturn with a value of 60280\,km. \\
Hence,
\begin {linenomath}
\begin{equation}
F_{\textnormal{CF}}\Delta \leq 3\cdot10^{-9}\left(\frac{{B_0}^2}{\mu_0}\right)\left(\frac{\Delta}{R_S}\right)
\end{equation}
\end {linenomath}
as, in reality, the density and rotation rate of the layer will be intermediate between those either side of the magnetopause. Also from \citet{Achilleos2010a} Figure 10 (middle panel),
\begin {linenomath}
\begin{equation}
\left(\frac{B^2}{2\mu_0} + P_{\textnormal{MIMI}}\right) \sim 0.02\left(\frac{{B_0}^2}{\mu_0}\right)
\end{equation}
\end {linenomath}
thus,
\begin {linenomath}
\begin{equation}
\frac{F_{\textnormal{CF}}\Delta}{\left(\frac{B^2}{2\mu_0} + P_{\textnormal{MIMI}}\right)} \leq \left(\frac{3\cdot10^{-9}}{0.02}\right)\left(\frac{\Delta}{R_\textnormal{S}}\right) \leq 10^{-7}\left(\frac{\Delta}{R_\textnormal{S}}\right)
\end{equation}
\end {linenomath}
$\Delta$ is of the order 1\,$R_\textnormal{S}$ \citep{Masters2011a}. Thus,
\begin {linenomath}
\begin{equation}
\frac{F_{\textnormal{CF}}\Delta}{\left(\frac{B^2}{2\mu_0} + P_{\textnormal{MIMI}}\right)} \leq 10^{-7}
\end{equation}
\end {linenomath}
hence the centrifugal force is very small compared to the magnetic and suprathermal plasma pressure gradients, and can safely be neglected.

\section{Magnetopause Crossing Observations}
\label{sec:data}
\begin{figure}
\noindent\includegraphics[width=0.9\textwidth, keepaspectratio]{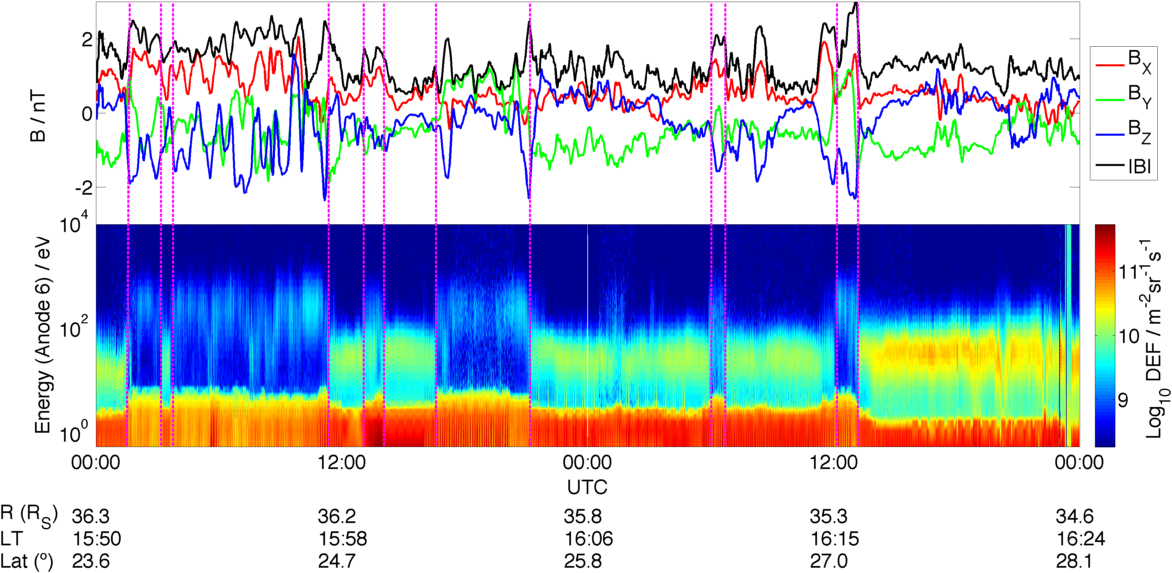} \\
\caption{In-situ data taken by the \textit{Cassini} spacecraft over 48 hours starting at midnight on day 123 of 2007 (3 May). The upper panel shows the components of the magnetic field as well as its magnitude with a one-minute time resolution smoothed using a moving average filter with a span of 11 minutes`. The lower panel shows an electron spectrogram from the CAPS-ELS instrument. The energy and count rate (which is proportional to density) of the electrons are represented logarithmically. The vertical magenta lines indicate magnetopause crossings, 12 such crossings can be seen during this period characterised by sudden changes in electron energy and count rate and magnetic field strength, as well as rotations in the magnetic field. The persistent population of low energy ($<$ 10\,eV) electrons are photoelectrons. }
\label{fig:caps_mag_data}
\end{figure}

\begin{figure}
\centering
\begin{minipage}{.49\textwidth}
  \centering
  \includegraphics[width=20pc]{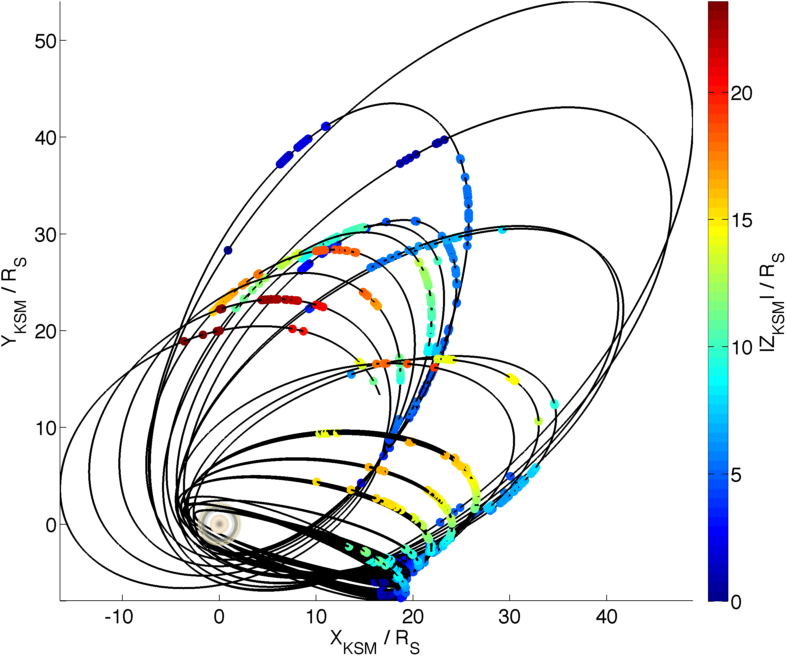}
  (a) Slice through the X-Y plane
\end{minipage}
\begin{minipage}{.49\textwidth}
  \centering
  \includegraphics[width=20pc]{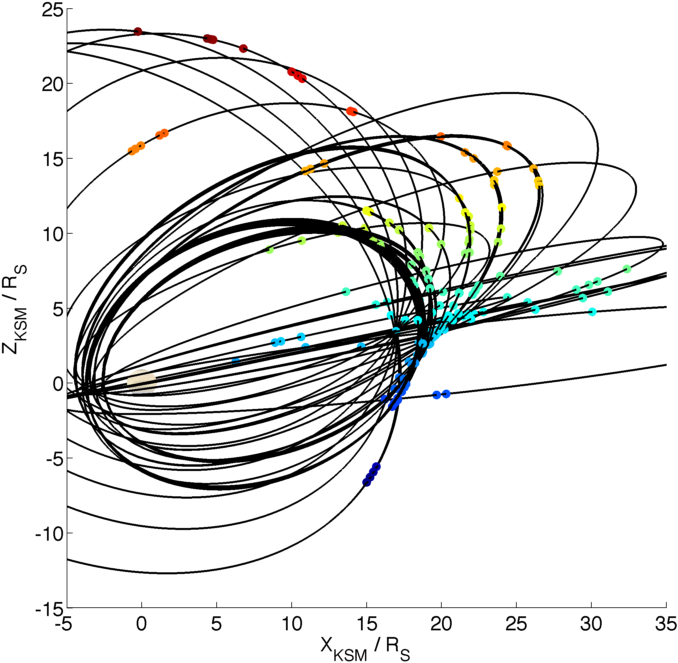}
  (b) Slice through the X-Z plane
\end{minipage}
\caption{The lines show consecutive orbits of the \textit{Cassini} spacecraft and the points are locations where a magnetopause crossing has been identified. Each crossing is coloured according to its $Z_{\textnormal{KSM}}$ coordinate. (a) is looking from the dawn side to the dusk side of the magnetosphere. It shows the high-latitude coverage of the spacecraft within the time period that this study covers. (b) is looking down on the magnetosphere from the pole of the northern hemisphere. It shows that the majority of the crossings are confined to the noon-dusk sector.}
\label{fig:crossing_locations}
\end{figure}

For this study, we examined magnetometer data from the beginning of 2007 up until the very end of 2008 during which \textit{Cassini} executed its first family of high-inclination orbits. Positive magnetopause crossing identifications were made in both the MAG and the CAPS-ELS datasets for each crossing used in the analysis.\\
Generally speaking, upon a transition from the magnetosheath to the magnetosphere, an increase in the total field strength as well as a rotation in the field is observed in the MAG data. In addition, the magnetic field is usually much steadier inside the magnetosphere. However, this is not always the case and a positive identification of a magnetopause crossing is not always possible using MAG data alone. In this situation, plasma data can provide complimentary information. \\
A sudden drop in the density of the plasma is observed in the CAPS-ELS data, typically by an order of magnitude, when the spacecraft passes from the magnetosheath to the magnetosphere. Also, the modal energy of the plasma just inside the magnetosphere is typically an order of magnitude greater than in the magnetosheath. As such, it is often much easier to detect magnetopause crossings in the CAPS-ELS data than the MAG data. A series of magnetopause crossings are shown in Figure \ref{fig:caps_mag_data} which highlight some of these tendencies. \\
\citet{Kanani2010} also included electron pressures derived from the CAPS-ELS instrument in their model, but found that the partial pressures of these electrons were on average 1-2 orders of magnitude smaller than those associated with the magnetic field and the suprathermal ions. This result allows us to neglect the electron pressure for the purposes of this study. \\
Through careful analysis of the field and plasma data, 626 magnetopause crossings were identified. Crossings within one hour of each other were averaged together, as it is likely that these were caused by boundary waves in the magnetopause surface. Temporal averaging of magnetopause crossings has been carried out in various forms by e.g. \citet{Slavin1981, Slavin1983, Slavin1985a, Huddleston1998}. \citet{Arridge2006} decided instead to average the crossings spatially to account for the different spacecraft velocities of \textit{Cassini} and \textit{Voyager}. \citet{Masters2012b} found that boundary waves have a period of $\sim$3 hours on the dusk side of the planet, as such we have verified that the results of this study are insensitive to averaging together crossings within this length of time. \\
It is important to note that not all of the crossings have reliable suprathermal plasma pressure moments. The hot plasma pressure moments, determined using MIMI, generally have a resolution of 10 minutes, the minimum window usually required in order to have reliable statistics for the computation of the moment. They are very variable in nature and an increase by an order of magnitude from one 10 minute-average to the next is not uncommon. The population contributing this pressure consists mainly of protons and ions of oxygen (dominated by $\textnormal{O}^+$). As oxygen ions are much more massive than protons, they contribute about four times more to the total pressure than a proton of similar velocity. As a result, it only takes a short interaction with a stream of these ions to increase the total pressure by an order of magnitude (N. Sergis, private communication).\\
Further statistical measures have been used in order to reduce the effect of this on our results. For each magnetopause crossing, the average magnetic field is taken inside the magnetosphere as close to the crossing as possible, over a representative interval of time. This interval must be long enough such that at least three measurements of the hot plasma pressure moments are within it. The relative difference between the median and the upper and lower quartiles of the hot plasma pressure moment is then found; this difference must be within a tolerance of 0.60 for the crossing to be accepted. The total number of crossings for which there were reliable pressure moments available was 196.\\
The crossing locations are shown in Figure \ref{fig:crossing_locations}. The Kronocentric Solar Magnetospheric (KSM) co-ordinate system has been used, in which the \textit{X}-axis is directed from the planet to the Sun and the \textit{Z}-axis is such that the magnetic dipole axis of the planet lies within the \textit{X}-\textit{Z} plane. The \textit{Y}-axis completes the right-handed set and is thus pointed towards dusk local time.\\
Much scatter can be seen in the crossing positions. This is because the observations have been made over $\sim$650 days and the magnetopause has thus experienced large variations in shape and size, largely in response to variations in the solar wind dynamic pressure.

\section{Magnetopause Modeling}
\label{sec:results}
\subsection{Initial Results}
\label{sec:initial_results}
\begin{figure}
\centering
\begin{minipage}{.49\textwidth}
  \centering
  \includegraphics[width=20pc]{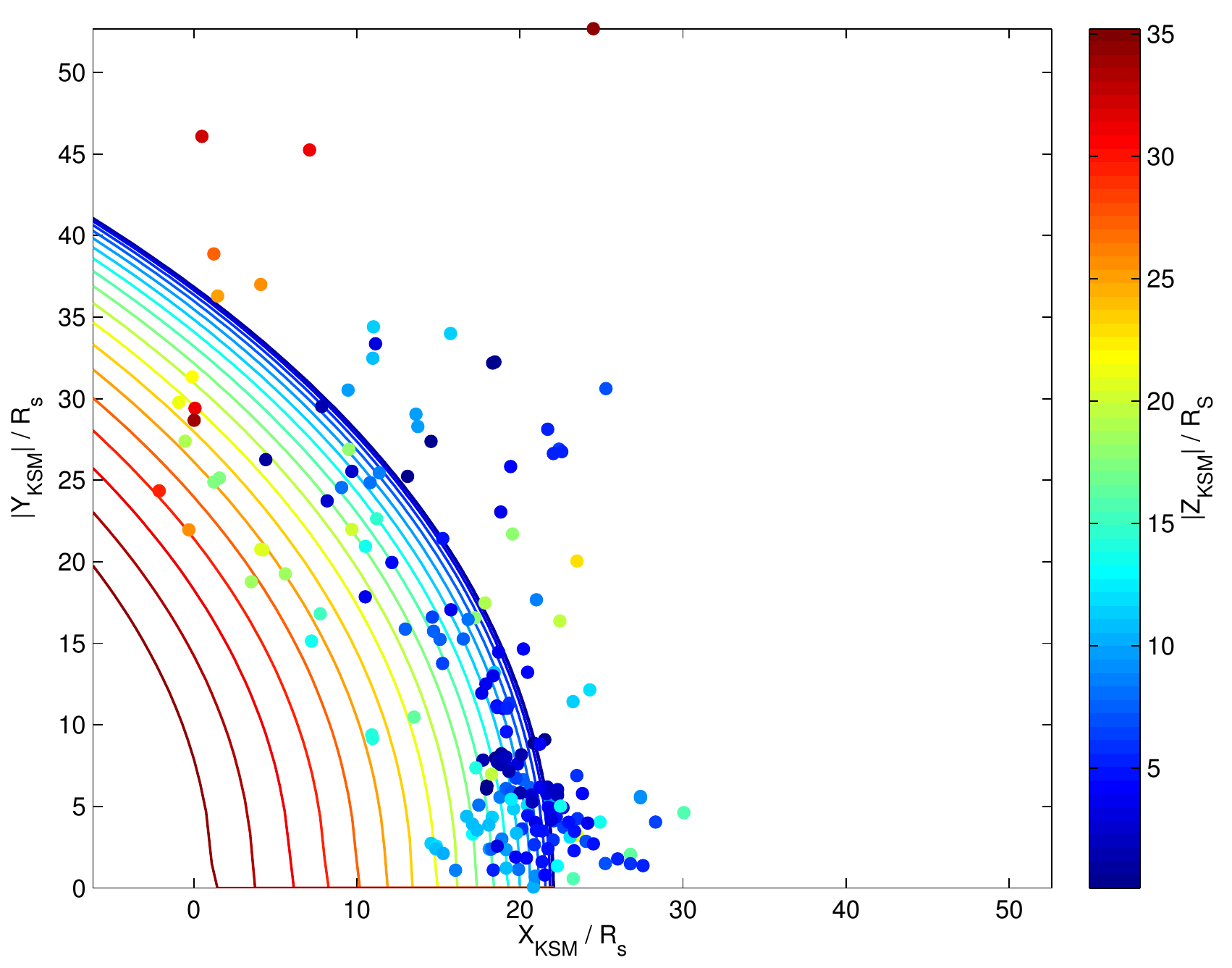}
  (a) Axisymmetric
\end{minipage}
\begin{minipage}{.49\textwidth}
  \centering
  \includegraphics[width=20pc]{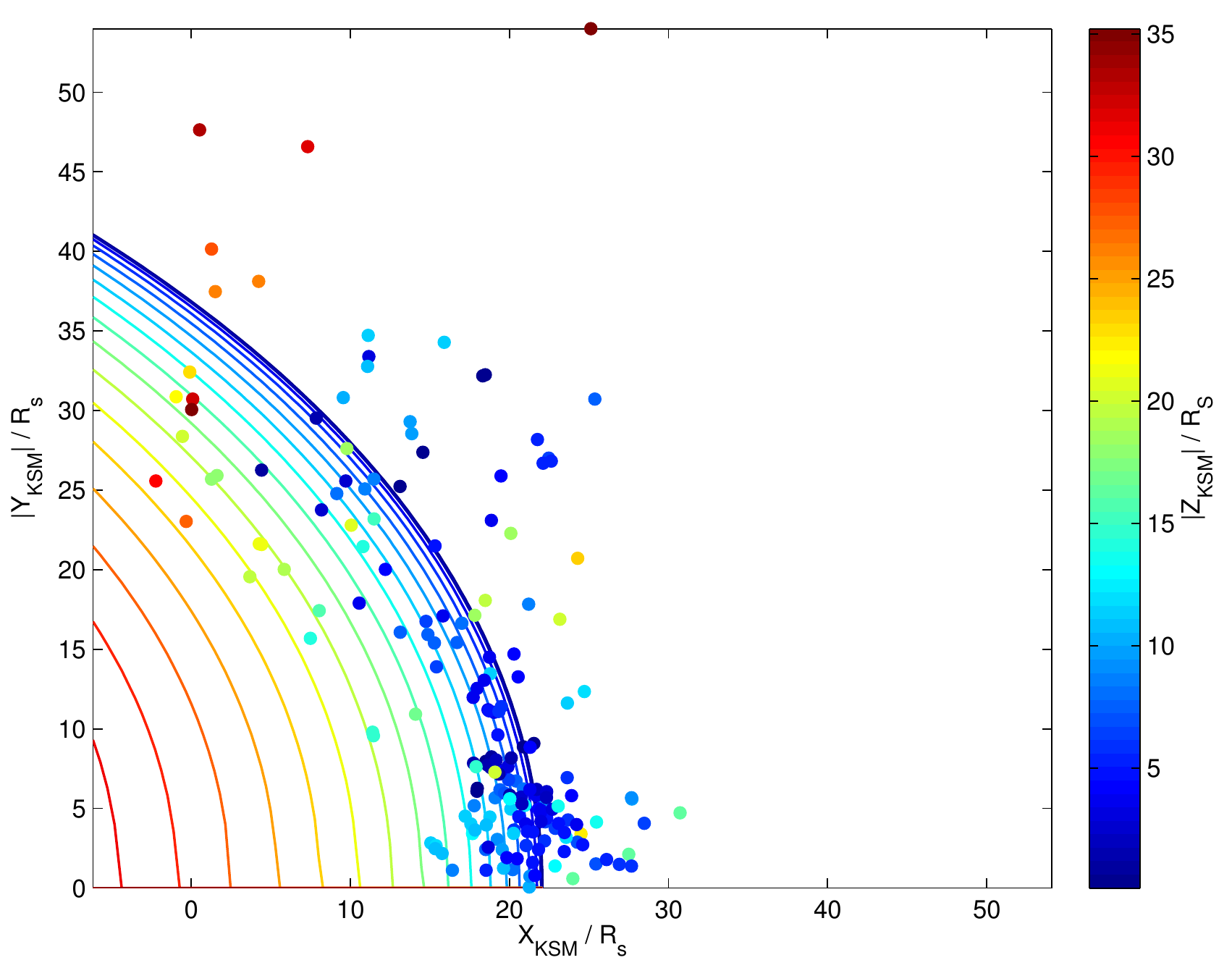}
  (b) Flattened by 20\%
\end{minipage}
\caption{Magnetopause crossing locations are scaled to the average solar wind dynamic pressure and are coloured by their $Z_{\textnormal{KSM}}$ coordinate. On top of this has been plotted an X-Y slice every 2\,$R_\text{S}$ from (a) the axisymmetric model of \protect\citet{Kanani2010} and (b) a flattened version of this model with a value of $\mathcal{E}$ of 0.80.}
\label{fig:x-y_slice}
\end{figure}

For present purposes, it is firstly necessary to normalise the crossing locations to predict where the magnetopause boundary would be located at a fixed value of solar wind dynamic pressure. More specifically,
\begin {linenomath}
\begin{equation}
\label{eq:normalise}
(X, Y, Z) = (X, Y, Z)_{\textnormal{OBS}}\left(\frac{D_P}{\langle D_P \rangle}\right)^{\frac{1}{\alpha}}
\end{equation}
\end {linenomath}
where $(X, Y, Z)$ and $(X, Y, Z)_{\textnormal{OBS}}$ are the scaled and observed coordinates of the crossing location and $\langle D_\textnormal{P} \rangle$ is a fixed pressure; the average dynamic pressure over all crossings was used in this case. Similar equations can be constructed for the other coordinates, as well as the radial distance from the planet centre. Although we normalise the crossings by $D_\textnormal{P}$ alone in this study, in reality the magnetic oscillation also affects the position of the magnetopause boundary. Here we assume that $D_\textnormal{P}$ is the dominant effect and neglect the effect of the magnetic oscillation, later in Section \ref{sec:skr} we will present evidence to support this assumption. \\
The normalised crossings are shown in Figure \ref{fig:x-y_slice} (a) along with representative X-Y slices of the axisymmetric model derived by \citet{Kanani2010} to form a contour map of the magnetosphere looking down onto the northern hemisphere. Although there is a degree of scatter (which will be discussed later in Section \ref{sec:reducing}), the crossings at $Z_{\textnormal{KSM}}$ {\lower 2pt \hbox{$\, \buildrel {\scriptstyle <}\over {\scriptstyle
\sim}\,$}} 10\,$R_\textnormal{S}$ tend to fit the slices well within an uncertainty of $1-2\,R_\textnormal{S}$. \\
However, as we increase $Z_{\textnormal{KSM}}$ (going from cooler colours to warmer colours in Figure \ref{fig:x-y_slice} (a)) the crossings seem to be systematically shifted away from the slices. This is particularly apparent for the crossings coloured from cyan-yellow on the colour scale, corresponding to 12-25\,$R_\textnormal{S}$ $Z_{\textnormal{KSM}}$. \\
A simple dilation of the magnetopause boundary along the $Z_{\textnormal{KSM}}$ direction has been used to construct a flattened magnetopause whereby $Z_{\textnormal{KSM}}$ in the axisymmetric boundary model is replaced by $\mathcal{E}Z_{\textnormal{KSM}}$. The factor $\mathcal{E}$ governs the degree of flattening; a value smaller (greater) than 1 represents a flattened (inflated) magnetosphere. \\
In Figure \ref{fig:x-y_slice} (b) a flattening of 20\% (i.e. $\mathcal{E}$ is equal to 0.80) has been applied to the model magnetopause surface. The positions of the crossings have changed somewhat, since the angle $\Psi$ is found by fitting the magnetopause surface through each crossing as done by \citet{Arridge2006} (see Appendix A3 of that paper). Since the surface is now flattened, the angular separation between the normal to the surface and the $X_{\textnormal{KSM}}$ axis is now larger and therefore $\Psi$ will be larger. This will result in a different estimate for $D_\textnormal{P}$ for each crossing following from Equation \ref{eq:Kanani}. This effect will be more pronounced for crossings at larger values of $Z_{\textnormal{KSM}}$ where the geometry of the magnetopause varies by the largest amount when this modification is applied. \\
In addition to this effect, some crossings present in the unflattened case may no longer be present when the magnetopause geometry is modified. This is because the Newton-Raphson iteration method used to fit the surface to each individual crossing cannot, in some cases, converge. The further the magnetopause is perturbed from the axisymmetric case by varying $\mathcal{E}$, the more difficult convergence seems to become. \\
There are no significant changes in the positions of the crossings between Figure \ref{fig:x-y_slice}. However, the crossings and their corresponding X-Y slices are much closer together now for the crossings at large $Z_{\textnormal{KSM}}$. This indicates that a surface flattened in the north-south direction is a better description for the magnetopause of Saturn, at least over the period at which these crossings were observed. \\
However, it is important to note that it is difficult to determine if the observed confinement is polar flattening, arising due to the disk-like nature of the obstacle to the solar wind flow, or if it was caused by seasonal effects related to the hinging of the magnetosphere. Saturn was approaching vernal (spring) equinox at the time that these observations were made, its magnetic dipole was tilted away from the Sun by an angle ranging from \hbox{$\sim$$13^\circ$ - $4^\circ$}. As a result, seasonal effects may play some role in confining the magnetosphere, particularly for the crossings near the start of the observation period. \\
The crossings at the largest values of $Z_{\textnormal{KSM}}$ (coloured orange and red in Figure \ref{fig:x-y_slice}) are of great interest to this study. These unusual crossings are very far from the \textit{X}-\textit{Y} slices of the same colour and clearly do not fit either model. \\
The plasma beta measured by the spacecraft for these crossings ranged between 15 and 25 and was exceptional compared to previous studies by \citet{Sergis2007, Sergis2009} and \citet{Masters2012a}, who found that the plasma beta just inside the magnetopause can be of the order 10. \\
It is possible that some form of transient event occurred during this time and was responsible for energising the plasma or contributing more suprathermal plasma to the system for this set of unusual crossings. In order to determine if a solar event may have caused this energisation, the solar wind dynamic pressure estimated using the procedure outlined above was compared to estimates from the Michigan Solar Wind Model (mSWiM) of \citet{Zieger2008}. This model uses data taken from many near-Earth spacecraft as an input and uses a one-dimensional MHD model to propagate this throughout the solar system as far as 10\,Au. \\
Its predictions are most accurate at near-apparent opposition between the Earth and the object of interest and is reasonably accurate within 75 days of apparent opposition. For Saturn, apparent opposition was at day 70 (11 March) 2007 which coincides with the beginning of the data set we have chosen to analyse. \\
The crossings of interest are hence within the time period where the model is accurate. However, the $D_\textnormal{P}$ predicted from our observations is approximately 60 times larger on average than that predicted by mSWiM which implies that these high plasma betas were not caused by a solar event. \\
The elevated plasma pressures observed for these crossings may be the result of ion conics as described by \citet{Mitchell2009}; these can remain relatively steady over a period of an hour or more. Alternatively, the magnetopause boundary may have simply been far from equilibrium at the time that the spacecraft crossed it. \\
As these observations were made over a period of 4 days on different trajectories but in similar regions of space, it is suggested that they could be related to a cusp region similar to that suggested by the modelling work of \citet{Maurice1996}.  \citet{McAndrews2008} observed magnetic field and plasma signatures suggestive of reconnection events at Saturn and found evidence of plasma energisation as a result. Since the cusp region is associated with newly reconnected field lines, this energisation could be the explanation for the large plasma beta values associated with these unusual crossings.

\subsection{Monitoring Pressure Equilibrium at the Magnetopause}
\label{sec:reducing}
\begin{figure}
\noindent\includegraphics[width=20pc]{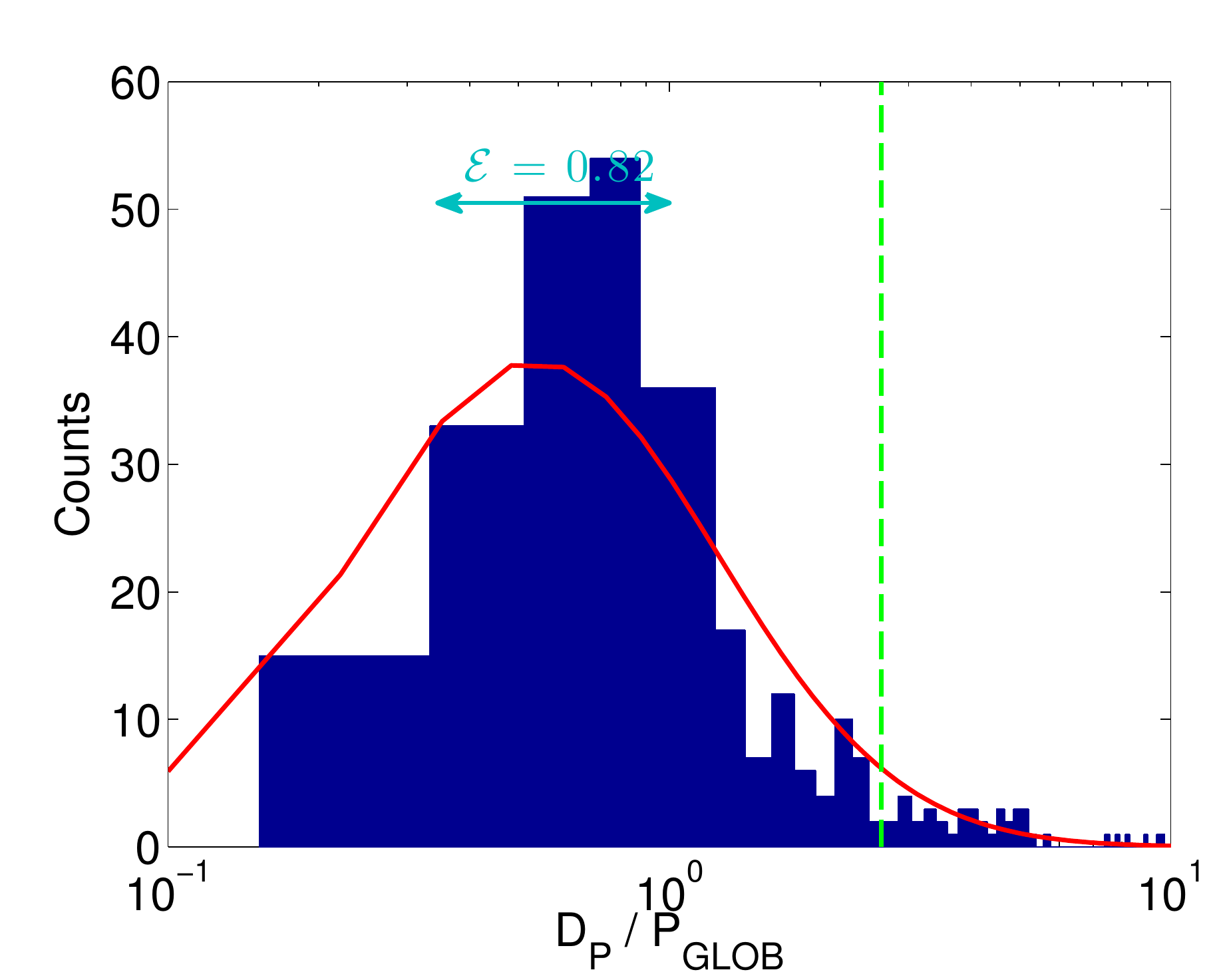}
\caption{The ratio of the pressure estimate found assuming pressure balance and the pressure estimate found by fitting the \protect\citet{Kanani2010} model through each crossing location has been calculated. This histogram shows the distribution of these ratios on a logarithmic axis and on top of this is plotted a log-normal curve (red line). The arrowed line indicates the range of ratios expected if an axisymmetric model is fitted through the surface of a flattened magnetopause model.}
\label{fig:hist_ratio}
\end{figure}

\begin{figure}
\centering
\begin{minipage}{.49\textwidth}
  \centering
  \includegraphics[width=20pc]{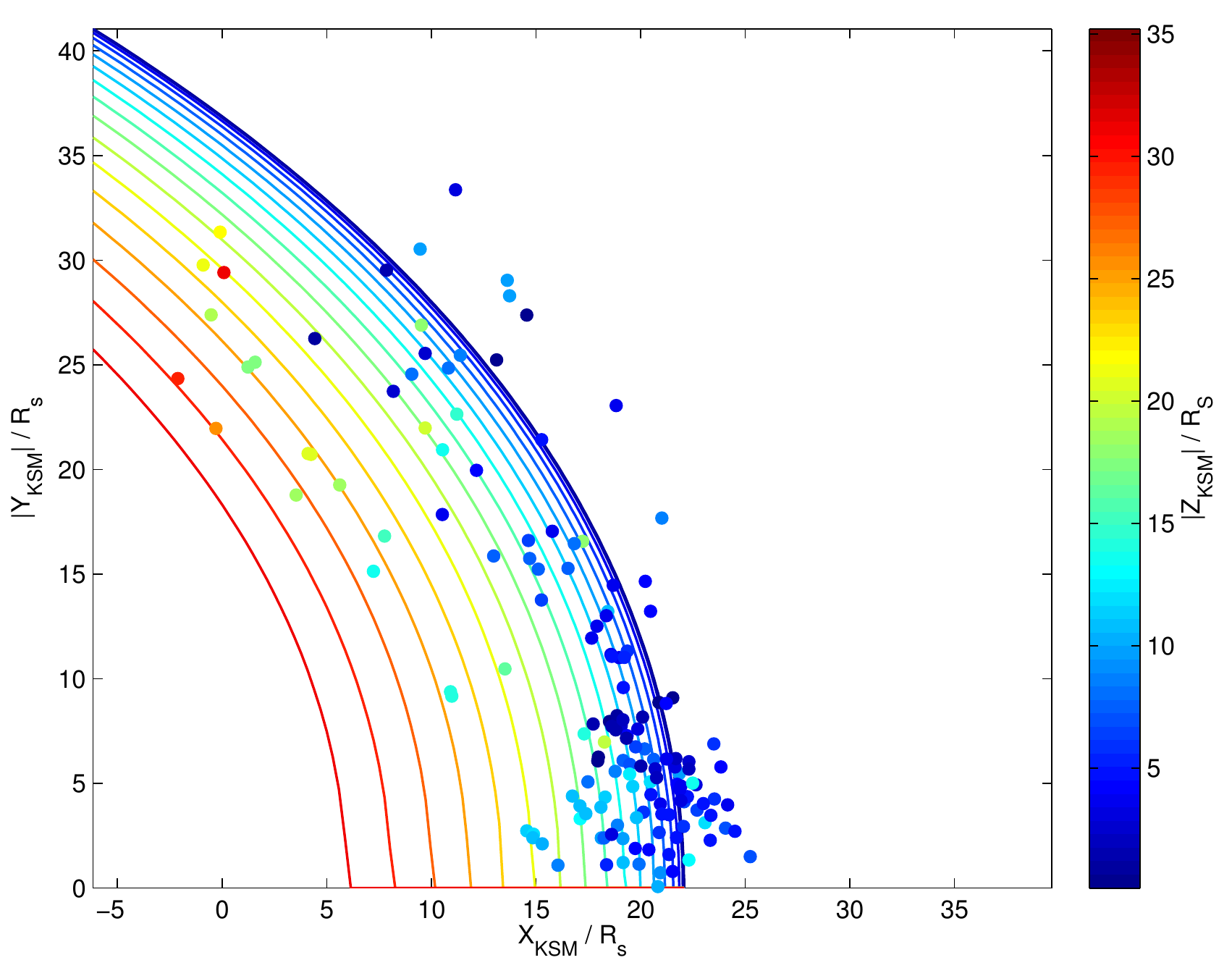}
  (a) Axisymmetric
\end{minipage}
\begin{minipage}{.49\textwidth}
  \centering
  \includegraphics[width=20pc]{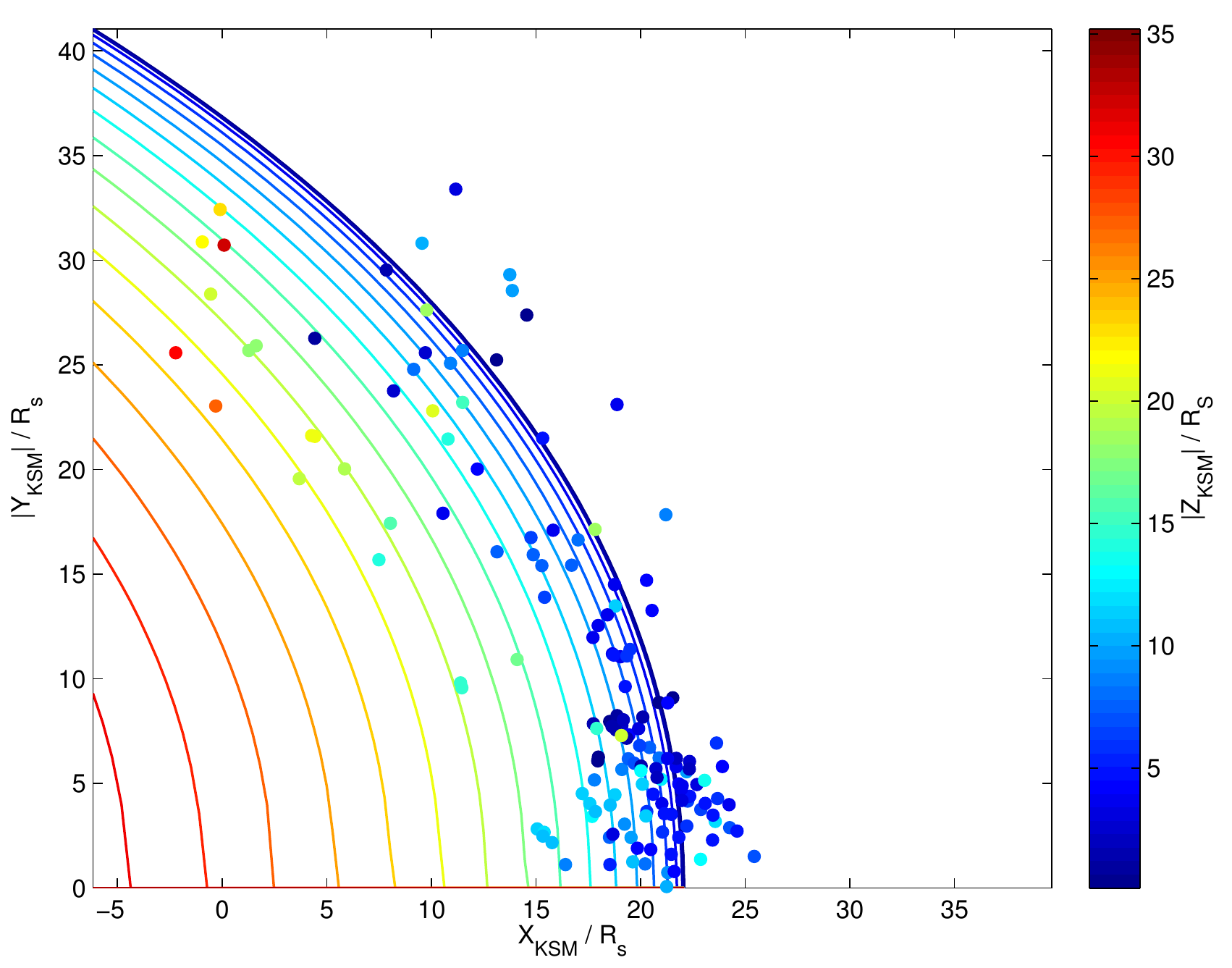}
  (b) Flattened by 20\%
\end{minipage}
\caption{Magnetopause crossing locations are scaled to the average solar wind dynamic pressure and are coloured by their $Z_{\textnormal{KSM}}$ coordinate. On top of this has been plotted X-Y slices every 2\,$R_\text{S}$ from (a) the axisymmetric model by \protect\citet{Kanani2010} and (b) a flattened version of this model with a value of $\mathcal{E}$ of 0.80. The data is filtered based on the pressure ratio shown in Figure \ref{fig:hist_ratio} to show crossings where the magnetopause can be assumed to be in equilibrium.}
\label{fig:x-y_slice_reduced}
\end{figure}

In accordance with the statistical law of large numbers, if a large number of measurements of its position are made over a sufficient period of time, the magnetopause is likely to be captured and depicted in an average state that is close to equilibrium. \\
Multi-spacecraft observations of the Earth's magnetopause by \citet{Dunlop2001} using the four \textit{Cluster} spacecraft found that strong and sudden accelerations of the magnetopause boundary can occur. As such, some scatter may exist in our data where static equilibrium of the boundary is not a good approximation. \\
Although we assumed that the magnetopause is in equilibrium at the outset of this study, we were able to quantify the departure from equilibrium as follows. A `global' pressure estimate was made for each magnetopause crossing by fitting the \citet{Kanani2010} model through each crossing and calculating the resulting pressure from the stand-off distance, $r_0$, using Equation \ref{eq:R0}. The ratio between $D_\textnormal{P}$ and this new pressure estimate, referred to as $P_{\textnormal{GLOB}}$, can then be calculated. \\
For an axisymmetric surface, the ratio $\frac{D_\textnormal{P}}{P_{\textnormal{GLOB}}}$ is expected to be unity at all points on the surface. When the surface departs from axisymmetry, the stand-off distance for an axisymmetric model fitted through any point on the surface where \hbox{$Z_{\textnormal{KSM}} \not= 0$} will be consistently underestimated compared to a flattened model, and hence $P_{\textnormal{GLOB}}$ will be consistently overestimated and the ratio will drop below unity. \\
This ratio was computed for each crossing and a histogram of the results plotted in Figure \ref{fig:hist_ratio}. On top of this, the range indicated by the arrowed line shows the range of values expected based on using an axisymmetric model to estimate $P_{\textnormal{GLOB}}$ for a surface which is actually flattened ($\mathcal{E}$ = 0.80). \\
The crossings follow a log-normal distribution which is largely enclosed by this interval but there exists some crossings that depart greatly from the expected range. These cannot be explained by the departure of the surface from axisymmetry and we suggest that they arise due to the departure of the surface from equilibrium, which is the underlying assumption made throughout this process. Hence, ratios that depart greatly from this range imply that the magnetopause was subject to strong accelerations at the time the observation was made. Values of $\frac{D_\textnormal{P}}{P_{\textnormal{GLOB}}}$ far from unity indicate a strong discrepancy between where \textit{Cassini} encountered the magnetopause and where it would have encountered it, had it been in equilibrium. As a result, this ratio can be used as a diagnostic to determine if the magnetopause was likely to be close to equilibrium when the spacecraft encountered it. \\
In Figure \ref{fig:x-y_slice_reduced} a new criterion has been included to filter the data set by removing any crossings where this ratio exceeds one interquartile range of the median (corresponding to an acceptance level of \hbox{$\sim2.4\sigma$}). This reduces the number of magnetopause crossings to 158, down from 196. It can be seen by comparing Figure \ref{fig:x-y_slice} to Figure \ref{fig:x-y_slice_reduced} that the amount of scatter in the normalised positions of the crossings has been reduced considerably through the application of this criterion.

\subsection{Statistical Tests}
\label{sec:stats}
\begin{figure}
\noindent\includegraphics[width=20pc]{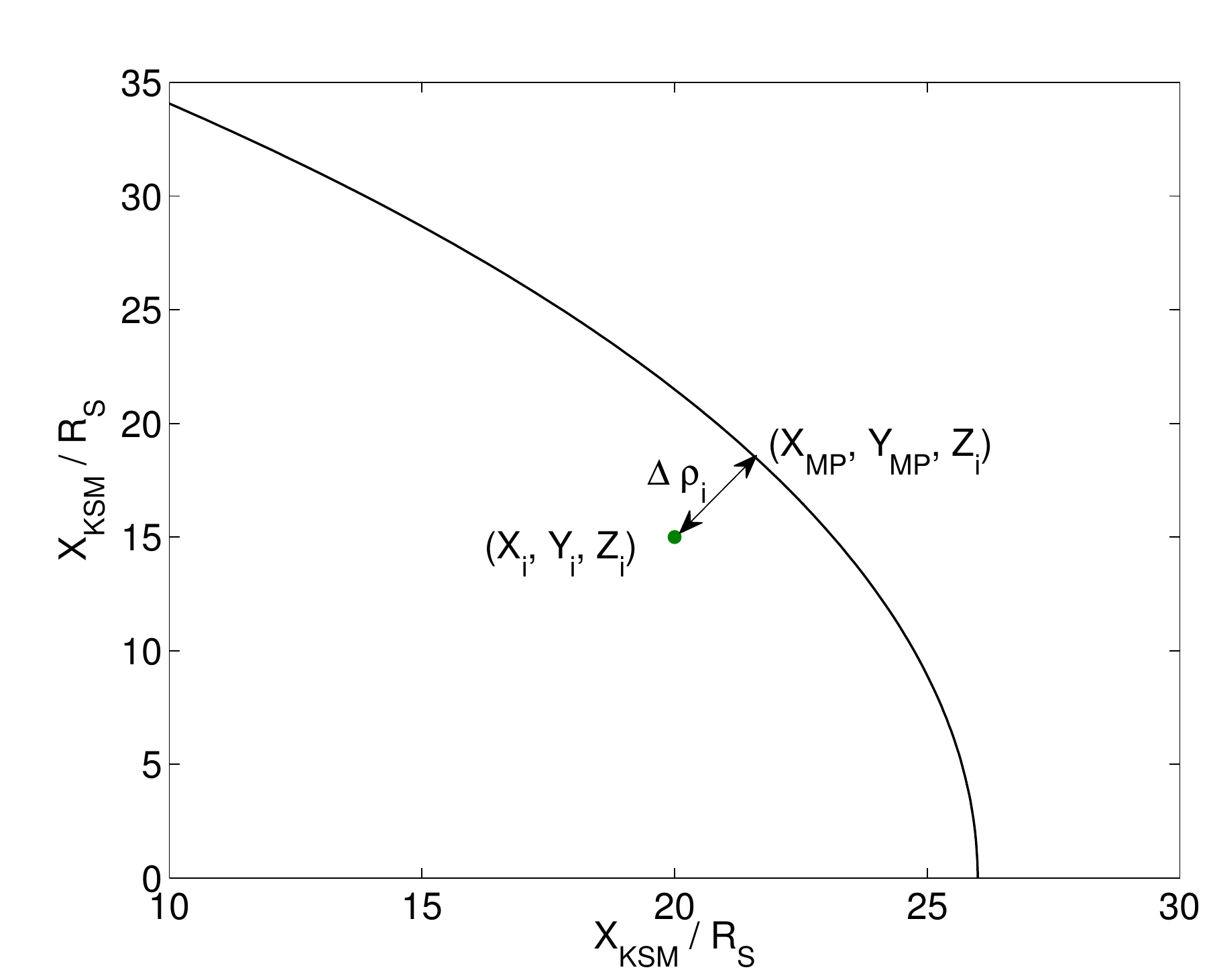}
\caption{Illustrated is the distance $\Delta\rho_i$, the distance between a normalised magnetopause crossing and an X-Y slice (at the $Z_{\textnormal{KSM}}$ coordinate of the crossing) from a magnetopause surface constructed at $\langle D_P \rangle$. $\langle D_P \rangle$ is the pressure to which the crossings are normalised.}
\label{fig:crossing_MP_dist}
\end{figure}

\begin{figure}
\centering
\begin{minipage}{.49\textwidth}
  \centering
  \includegraphics[width=20pc]{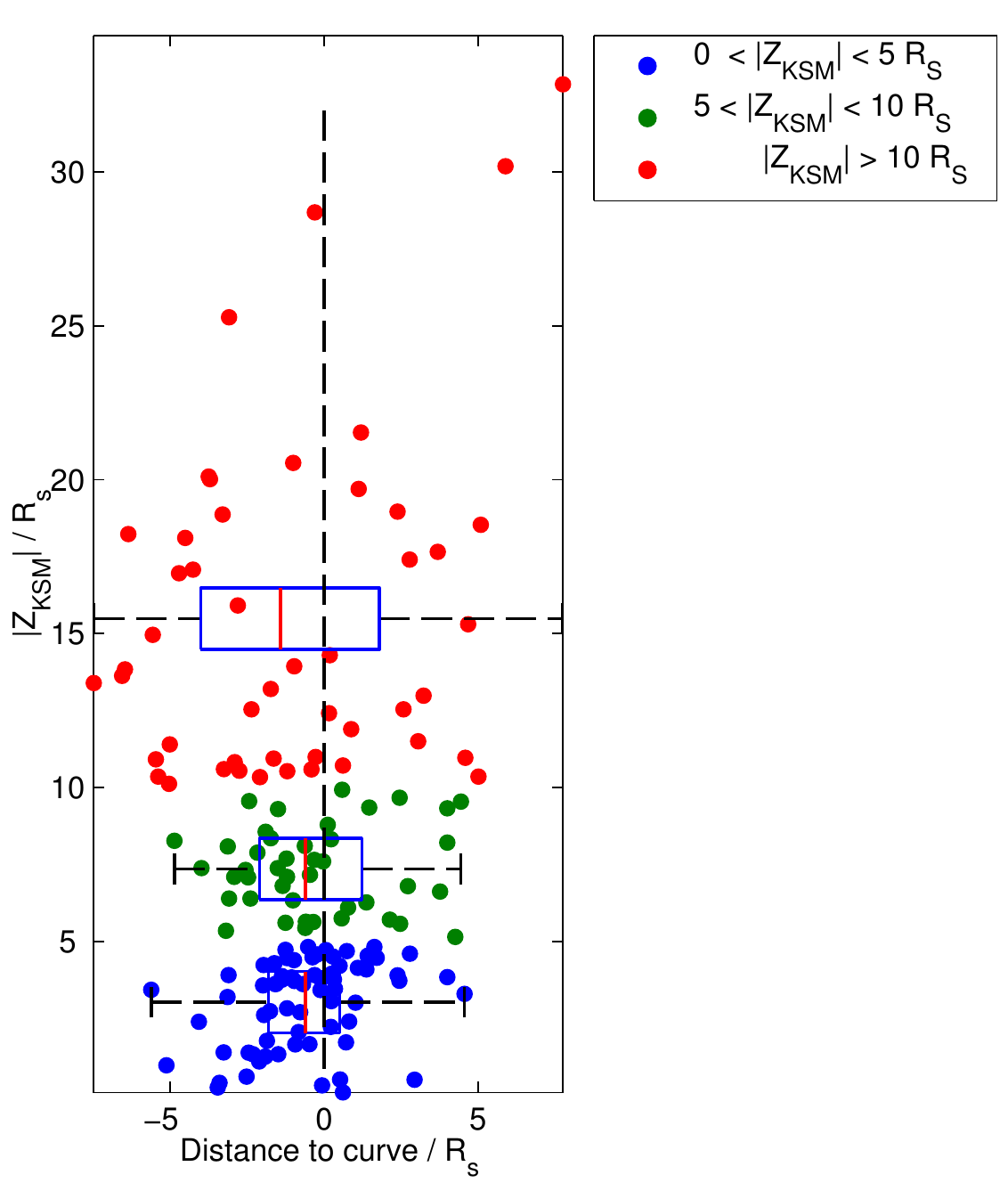}
  (a) Axisymmetric
\end{minipage}
\begin{minipage}{.49\textwidth}
  \centering
  \includegraphics[width=20pc]{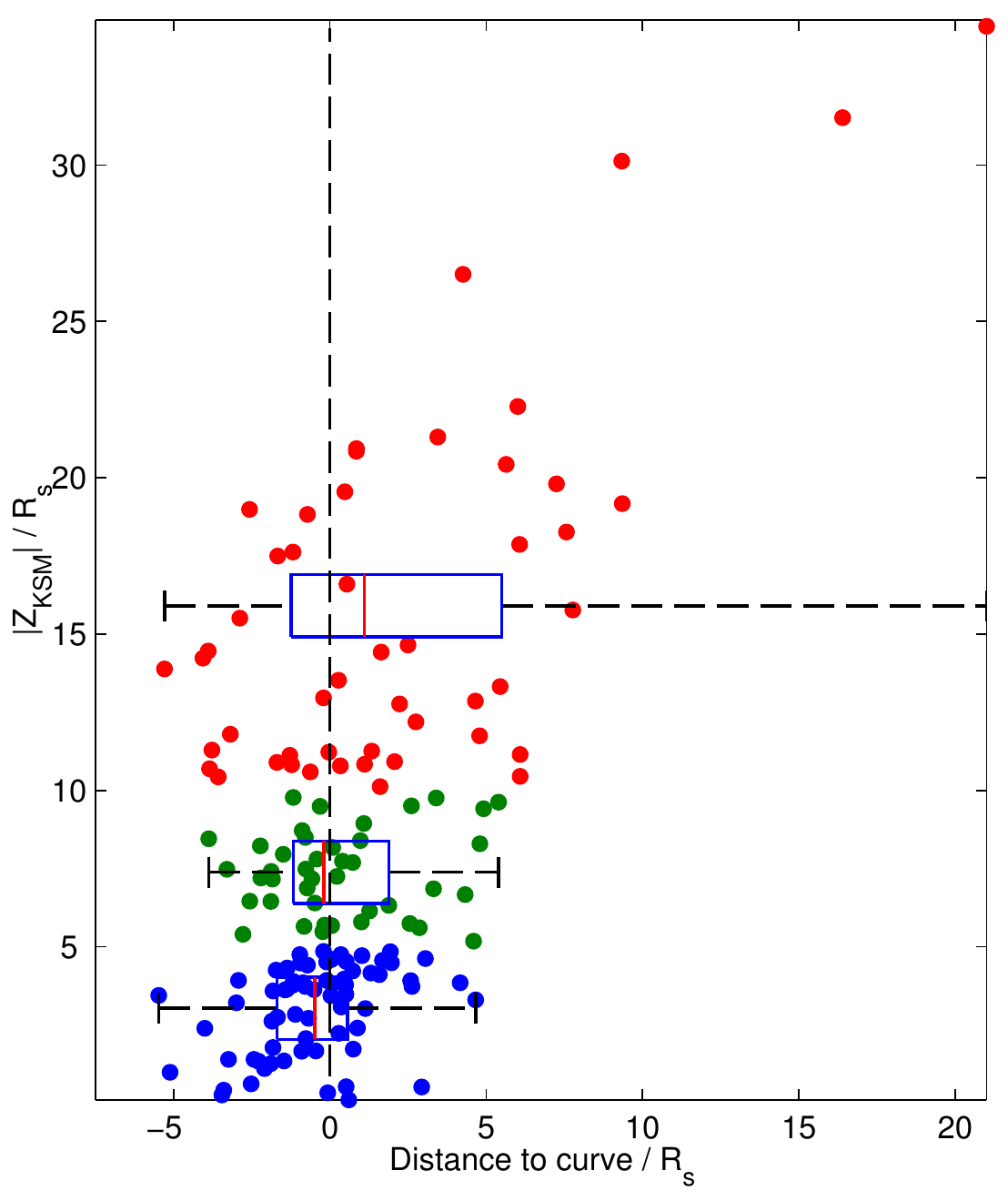}
  (b) Flattened by 20\%
\end{minipage}
\caption{Plotted is the distance between the normalised X-Y position of each crossing and the X-Y slice plotted at the $Z_{\textnormal{KSM}}$ coordinate of each crossing from (a) the axisymmetric model of \protect\citet{Kanani2010} and (b) a flattened version of this model with a value of $\mathcal{E}$ of 0.80. The distance has been defined such that if a crossing is above (below) the magnetopause the distance is positive (negative). The crossings have been arbitrarily separated into bands of $Z_{\textnormal{KSM}}$ at 5\,$R_\textnormal{S}$ intervals, but the minimum number of crossings within each band has been kept at 25 by collapsing the bands at large $Z_{\textnormal{KSM}}$. Adjusting the width of these bands has little effect on the results of the statistics shown in Table \ref{tb:stats}. A box plot has been plotted on top of this for each band. The reduced data set has been used.}
\label{fig:delta}
\end{figure}

Let us consider the Wilcoxon signed rank test \citep{Wilcoxon1945}, applied to a subsample of our magnetopause crossings which lie in a given range $\{ Z_0, Z_f \}$ of the $Z_{\textnormal{KSM}}$ coordinate. We will refer to this range as the `$Z_{\textnormal{KSM}}$ band' of this subsample. \\
The quantity, $\Delta \rho$ to which we apply the rank sum test are defined as follows:
\begin {linenomath}
\begin{equation}
\Delta \rho_i = [(X_{\textnormal{MP}}(Z_i, \mathcal{E}) - X_i)^2 + (Y_{\textnormal{MP}}(Z_i, \mathcal{E}) - Y_i)^2]^\frac{1}{2}
\end{equation}
\end {linenomath}
the symbols here have the following meanings. There are $N$ normalised crossings in the given $Z_{\textnormal{KSM}}$ band, and the $i$th crossing position is at \hbox{$(X_i, Y_i, Z_i)$} in KSM coordinates. The magnetopause surface is then constructed at \hbox{$\langle D_P \rangle$}, the pressure to which the crossings have been normalised, and the point on the slice through the X-Y plane at $Z_i$, closest to the normalised crossing location, has coordinates $(X_{MP}, Y_{MP}, Z_i)$. Hence $\Delta\rho_i$ represents a distance between each normalised crossing and the X-Y slice (at the same value of $Z_{\textnormal{KSM}}$) of a magnetopause model with a particular flattening parameter and is illustrated in Figure \ref{fig:crossing_MP_dist}. The two sets of such distances we wish to compare are simply the $\Delta\rho_i$ evaluated for \hbox{$\mathcal{E}$ = 1} (the axisymmetric case) and \hbox{$\mathcal{E}$ $<$ 1} (a flattened magnetopause model). \\
Let us represent these two sets of distances as \hbox{$\Delta\rho_i$($\mathcal{E}$ = 1)} and \hbox{$\Delta\rho_i$($\mathcal{E}$ = 0.80)} with \hbox{$i$ = 1 ... $N$}. As we shall see in the following analysis, \hbox{$\mathcal{E}$ = 0.80} appears to give values of $\Delta\rho_i$ significantly closer to 0 (i.e. a better agreement between modelled and observed magnetopause location) at high latitudes. \\
To apply the signed rank test, we order the union of both sets \hbox{$\Delta\rho_i$($\mathcal{E}$ = 1)} and \hbox{$\Delta\rho_i$($\mathcal{E}$ = 0.80)} in ascending order ignoring the signs, assigning a rank of 1 to the lowest absolute value, 2 to the second lowest absolute value, and so on. Each rank is then labelled with its sign according to the sign of $\Delta\rho_i$, $s_i$. If we denote these rank values as $rank(\Delta\rho_i)$, then the rank sums of the two distance sets are defined as $|\displaystyle\sum\limits_{i=1}^n s_i \cdot rank(\Delta\rho_i(\mathcal{E} = 1))|$ and $|\displaystyle\sum\limits_{i=1}^n s_i \cdot rank(\Delta\rho_i(\mathcal{E} = 0.80))|$. \\
The test statistic, $W$, is the rank sum of either group of $\Delta\rho_i$ values. The test itself consists of comparing $W$ with a critical value $S(p,N)$. This value $S(p,N)$ denotes that, if the the null hypothesis is true (i.e. the median values of the underlying $\Delta\rho_i$ distributions are zero), there is a probability $p$ that $W$ would exceed $S(p,N)$. Table \ref{tb:stats} indicates the value of $p$ for which $W$ = $S(p,N)$ for different $Z_{\textnormal{KSM}}$ bands. This tabulated value may be thought of as the probability that the magnetopause crossings would have a signed rank equal to or exceeding their observed value, given that they correspond to a magnetopause with a given flattening parameter $\mathcal{E}$. \\
The results of this test applied to the data shown in Figure \ref{fig:delta} are shown in Table \ref{tb:stats}. In each case, the \emph{p}-value when a flattening is applied to the model magnetopause is larger than that of the axisymmetric model, which indicates that, on average, flattening the magnetopause surface causes it to move closer to the crossing positions and hence a flattened surface is a better fit to the data. \\
This test is powerful as it doesn't rely on the population being normally distributed unlike similar tests, for example the Student's t-test. However, it does assume that the population is symmetric. The adjusted Fisher-Pearson standardised moment coefficient \citep{Doane2011} provides a measure of sample symmetry and has been calculated for each $Z_{\textnormal{KSM}}$ band to determine if it is, indeed, symmetric and, hence, if the Wilcoxon signed rank test applies. A coefficient of zero indicates that the data is perfectly symmetric, however this is very unlikely for real-world data. \citet{Doane2010} compiled a table of sample skewness coefficients corrected for sample size. We compare the results of this test of symmetry against these criteria to determine that each band of $Z_{\textnormal{KSM}}$ is approximately symmetric as the calculated coefficients are well within their limits. As such, we deem the Wilcoxon signed rank test to be appropriate for our data.

\subsection{Uncertainty in $\mathcal{E}$}
\label{sec:uncertainty}
It became clear that the value of $\mathcal{E}$ that provided the best fit between the model and the data is dependant on the distribution of the magnetopause crossings within the bands of $Z_{\textnormal{KSM}}$. Hence it is also dependant on the binning process itself. We have used this fact to estimate the uncertainty in our estimate of $\mathcal{E}$. \\
For each estimate of $\mathcal{E}$, the bands have been spaced evenly by $Z_{\textnormal{KSM}} = \Delta Z_{\textnormal{KSM}}$. To ensure that a significant number of crossings lie within each band, the minimum value chosen for $\Delta Z_{\textnormal{KSM}}$ was 5\,$R_\textnormal{S}$ and this was increased in steps of 1\,$R_\textnormal{S}$ up to a maximum value of 10\,$R_\textnormal{S}$. \\
As there are many more crossings at smaller values of $Z_{\textnormal{KSM}}$, it was found that bands at values of $Z_{\textnormal{KSM}}$ between 20-35\,$R_\textnormal{S}$ were sparsely populated. Crossings in these bands were moved down to the band below until a minimum of 25 crossings occupied each band since it is difficult to determine if a sample smaller than 25 is symmetric. This ensures that there are statistically enough crossings within each band to enable us to draw conclusions from the data. \\
This procedure yielded a mean, modal and median value of $\mathcal{E}$ of 0.82 with an uncertainty of $\pm$ 0.03 indicating a polar confinement of {20\% $\pm$ 3\%}. \\
A second uncertainty estimate has been made using a Monte Carlo (BCa bootstrap) method. Here, the same procedure used to determine $\mathcal{E}$ in Section \ref{sec:stats} is used, but this time \textit{N} crossings are randomly drawn with replacement from our population of \textit{N} crossings. As the crossings are drawn with replacement, a different set of crossings are drawn for each resampling and, as a result, the best fitting value of $\mathcal{E}$ changes. This was repeated 2000 times and  $\mathcal{E}$ was found to be 0.81 within a confidence interval of {0.75-0.84} at the 68.3\% (1\,$\sigma$) confidence level.

\subsection{Magnetopause Pressure Dependance}
\label{sec:highlowpressure}
\begin{figure}
\centering
\begin{minipage}{.49\textwidth}
  \centering
  \includegraphics[width=20pc]{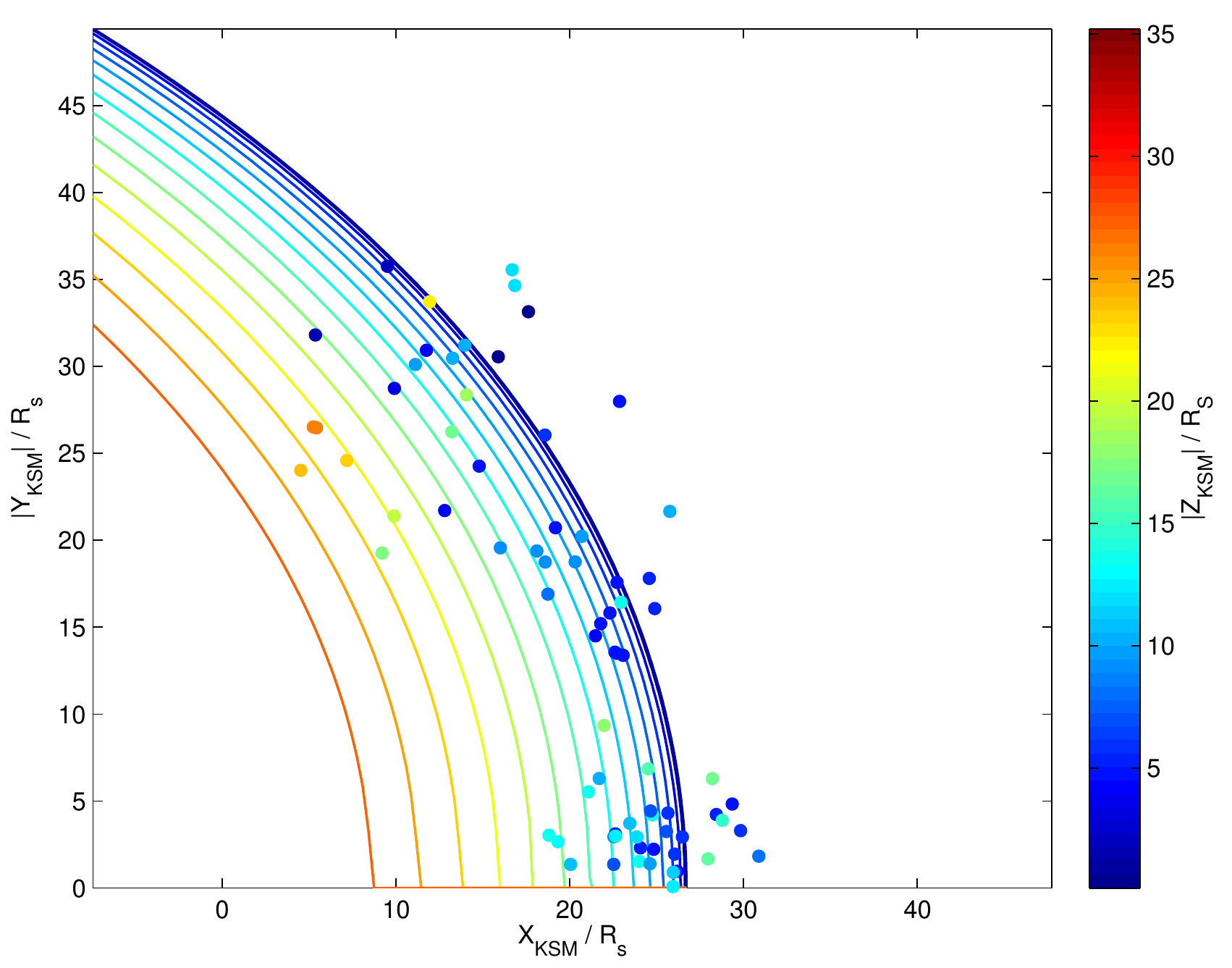}
  (a) Low Pressure Crossings
\end{minipage}
\begin{minipage}{.49\textwidth}
  \centering
  \includegraphics[width=20pc]{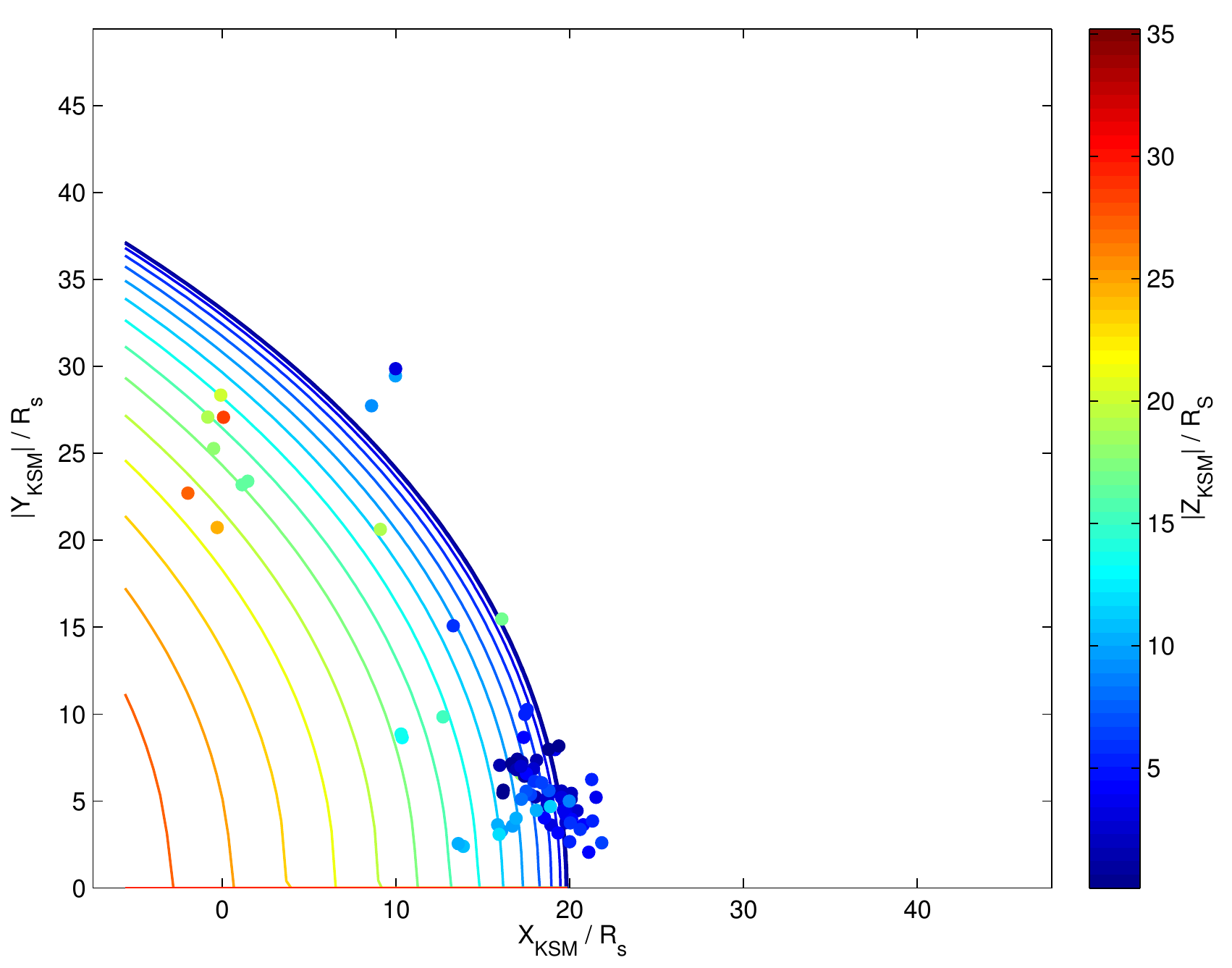}
  (b) High Pressure Crossings
\end{minipage}
\caption{The crossings have been split into two populations, (a) one of lower than average dynamic pressure and (b) one of higher than average dynamic pressure. X-Y slices of a model flattened using a value of $\mathcal{E}$ of 0.81 have then been plotted over the crossing positions.}
\label{fig:psw_high_low}
\end{figure}

The reduced set of crossings has been separated into two groups in Figure \ref{fig:psw_high_low}, one where the estimated dynamic pressure is above the average and where it is below. There seems to be a better distribution of crossings with low $D_\textnormal{P}$ whereas the crossings at high $D_\textnormal{P}$ seem to be clustered within 3 regions of local time. \\
The same techniques as previously discussed have been used to determine the value of $\mathcal{E}$ that provides the best fit between the model and each data sample. For the low $D_\textnormal{P}$ crossings, a value of $\mathcal{E}$ of 0.75 within a confidence interval of \hbox{0.71 - 0.82} provides the best fit whereas for the high $D_\textnormal{P}$ crossings a value of 0.79 within a confidence interval of \hbox{0.78 - 0.93} was found. Hence, within the given uncertainties and for the range of dynamic pressures considered here, the polar confinement of the magnetosphere is insensitive to changes in dynamic pressure. \\
The flaring parameter, $\textit{K}$, is the same in both cases within its uncertainty, based on the uncertainties in the model parameters determined by \citet{Kanani2010}. This contrasts with the study by \citet{Huddleston1998} who found the Jovian magnetosphere to be more streamlined at high $D_\textnormal{P}$. 

\subsection{Trajectory Analysis}
\label{sec:traj}
\begin{figure}
\centering
\begin{minipage}{.49\textwidth}
  \centering
  \includegraphics[width=20pc]{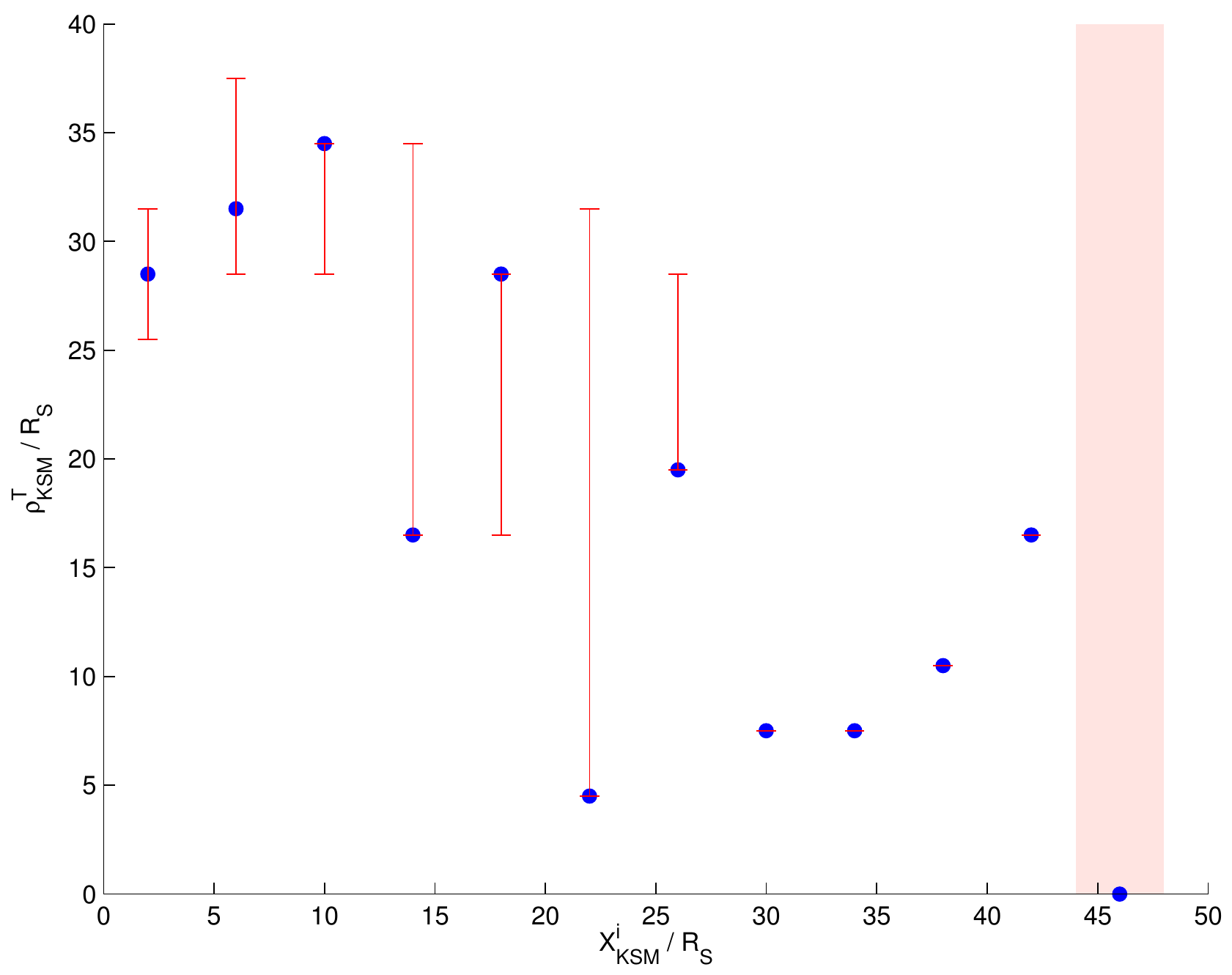}
  (a) Trajectory in Equatorial Regions
\end{minipage}
\begin{minipage}{.49\textwidth}
  \centering
  \includegraphics[width=20pc]{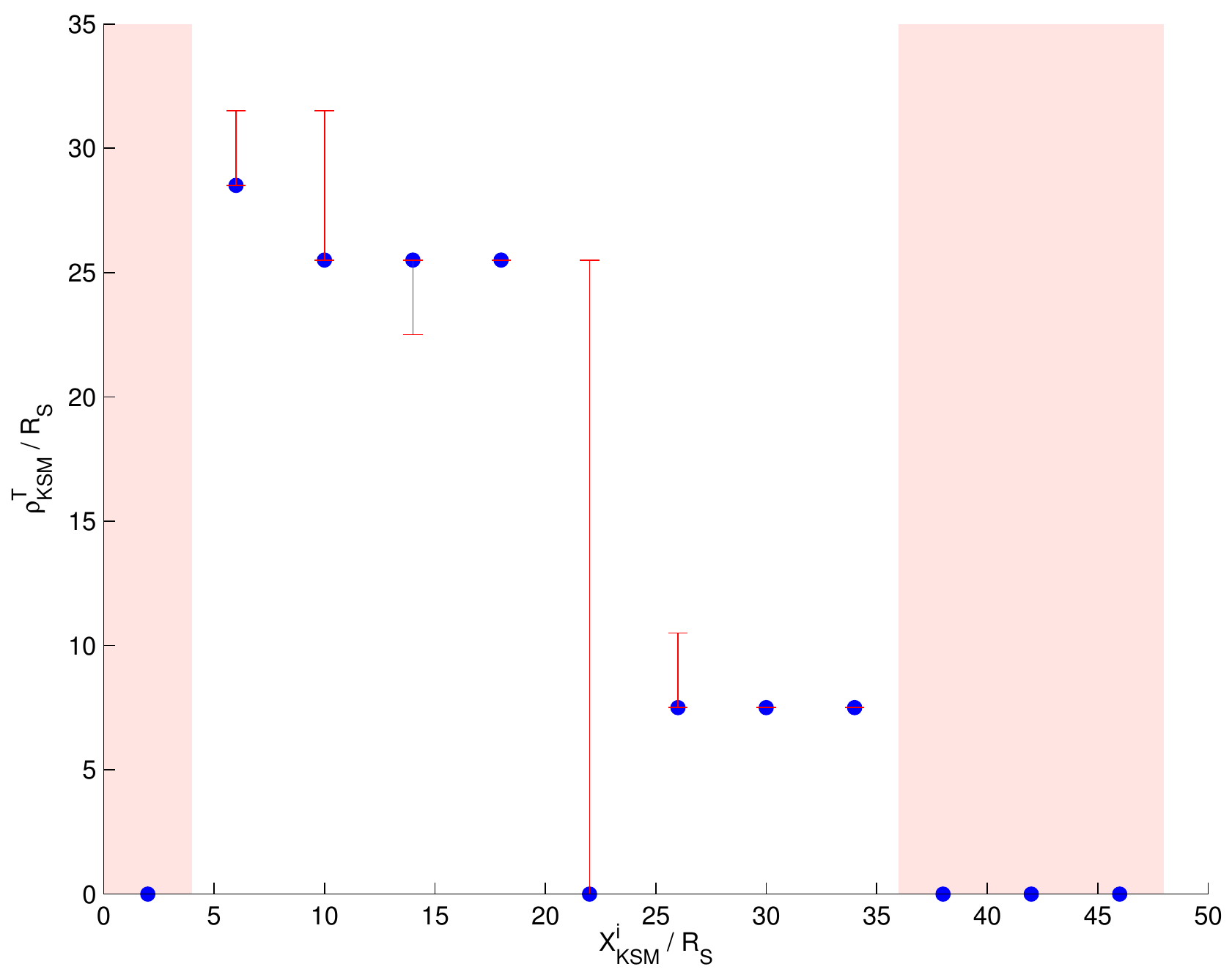}
  (b) Trajectory in Polar Regions
\end{minipage}
\caption{The `transition distance', $\rho^T_{\textnormal{KSM}}$, has been determined through analysis of the spacecraft trajectories over which the magnetopause crossings used in this study were found, and is plotted for each $X^i_{\textnormal{KSM}}$ bin as blue points. Regions where a transition distance could not be identified are shaded red. Confidence intervals are also included and were determined using Monte Carlo Bootstrap simulations resampling 1000 times for each $X^i_{\textnormal{KSM}}$ bin.}
\label{fig:traj_results}
\end{figure}

In order to be sure that we are adequately sampling the mean position of the boundary, we need to analyse the spacecraft trajectory and ensure that, even for the high-latitude passes, there is a clear transition between where we spend 50\% of the time inside, and 50\% outside, the magnetosphere. A similar method to that employed by \citet{Joy2002} in the case of Jupiter and subsequently employed by \citet{Achilleos2008} in the case of Saturn is used here. The procedure is as follows:\\
1. Magnetopause crossings are located and the spacecraft trajectory is split into small time intervals. It is important that the list of crossings is as complete as possible to obtain accurate results. It is also important that the sampling time scale is small compared to the time between crossings. An interval of 10 minutes is used here. \\
2. At each point along the magnetopause trajectory, the magnetopause crossings are used to determine if the spacecraft is inside or outside of the magnetosphere and \hbox{$\rho_{\textnormal{KSM}} = \sqrt{Y_{\textnormal{KSM}}^2 + Z_{\textnormal{KSM}}^2}$} and \hbox{$\phi_{\textnormal{KSM}} = \tan ^{ - 1}\left(\frac{Z_{\textnormal{KSM}}}{Y_{\textnormal{KSM}}}\right)$} are calculated at each point in time. Occasionally this may be impossible due to a data gap or other such anomaly, in which case this interval is discarded. \\
3. Separate the crossings into bins of $X_{\textnormal{KSM}}$ of width $\Delta X_{\textnormal{KSM}}$ such that there are $N_X$ bins with centres $X^i_{\textnormal{KSM}}$ for \hbox{$i = 1 ... N_X$}. Further subdivide bins $X^i_{\textnormal{KSM}}$ into bins of $\rho_{\textnormal{KSM}}$ of width $\Delta\rho_{\textnormal{KSM}}$ with centres $\rho^j_{\textnormal{KSM}}$ for \hbox{$j = 1 ... N_\rho$} such that there are now $N_XN_\rho$ bins each containing $M_\textnormal{$I_{ij}$}$ data points inside the magnetosphere and $M_\textnormal{$O_{ij}$}$ data points outside the magnetosphere. \\
4. For each $X^i_{\textnormal{KSM}}$ bin, calculate the probability distribution of $\rho_{\textnormal{KSM}}$. The probability that the actual value of $\rho_{\textnormal{KSM}}$ exceeds that of  $\rho^j_{\textnormal{KSM}}$ can be estimated as \hbox{$P(\rho_{\textnormal{KSM}} > \rho^j_{\textnormal{KSM}}) = M_\textnormal{$I_{ij}$} / (M_\textnormal{$I_{ij}$} + M_\textnormal{$O_{ij}$})$}. \\
5. For each $X^i_{\textnormal{KSM}}$ bin, identify if there is a clear `transition distance', $\rho^T_{\textnormal{KSM}}$, where the spacecraft spends 50\% of the time inside, and 50\% outside, the magnetosphere. If this is the case, the magnetopause is being adequately sampled by the trajectories along which the magnetopause crossings are identified for that particular bin of $X_{\textnormal{KSM}}$. \\
The data points were then separated into two groups based on their $\phi_{\textnormal{KSM}}$ angles into equatorial ($\phi_{\textnormal{KSM}} < 50^\circ$) and high-latitude ($\phi_{\textnormal{KSM}} > 50^\circ$) parts of the trajectory. These data limits are valid only for magnetopause crossings in the northern hemisphere/near-equatorial southern magnetopause crossings. \\
Figure \ref{fig:traj_results} shows the results of following this procedure for our dataset. Included are error bars determined using a Monte Carlo Bootstrap method ran 1000 times for each $X^i_{\textnormal{KSM}}$ bin at the $3\sigma$ (99.7\%) level. It shows that in most cases, the transition distance is captured and hence the magnetopause is being adequately sampled. Figure \ref{fig:traj_results} (b) shows the results for the high-latitude parts of the trajectory. Most of the high-latitude crossings that show a large degree of flattening were in the $X_{\textnormal{KSM}}$ = 0 - 25 $R_\textnormal{S}$ range as can be seen in Figure \ref{fig:crossing_locations} (b). We capture the transition distance in all but one of the $X^i_{\textnormal{KSM}}$ bins in this range. In addition to this, on average the transition distance is a few $R_\textnormal{S}$ smaller in the polar dataset than in the equatorial dataset, consistent with a polar confinement.

\subsection{Phase of the Global Magnetic Oscillation}
\label{sec:skr}
\begin{figure}
\centering
\begin{minipage}{.49\textwidth}
  \centering
  \includegraphics[width=20pc]{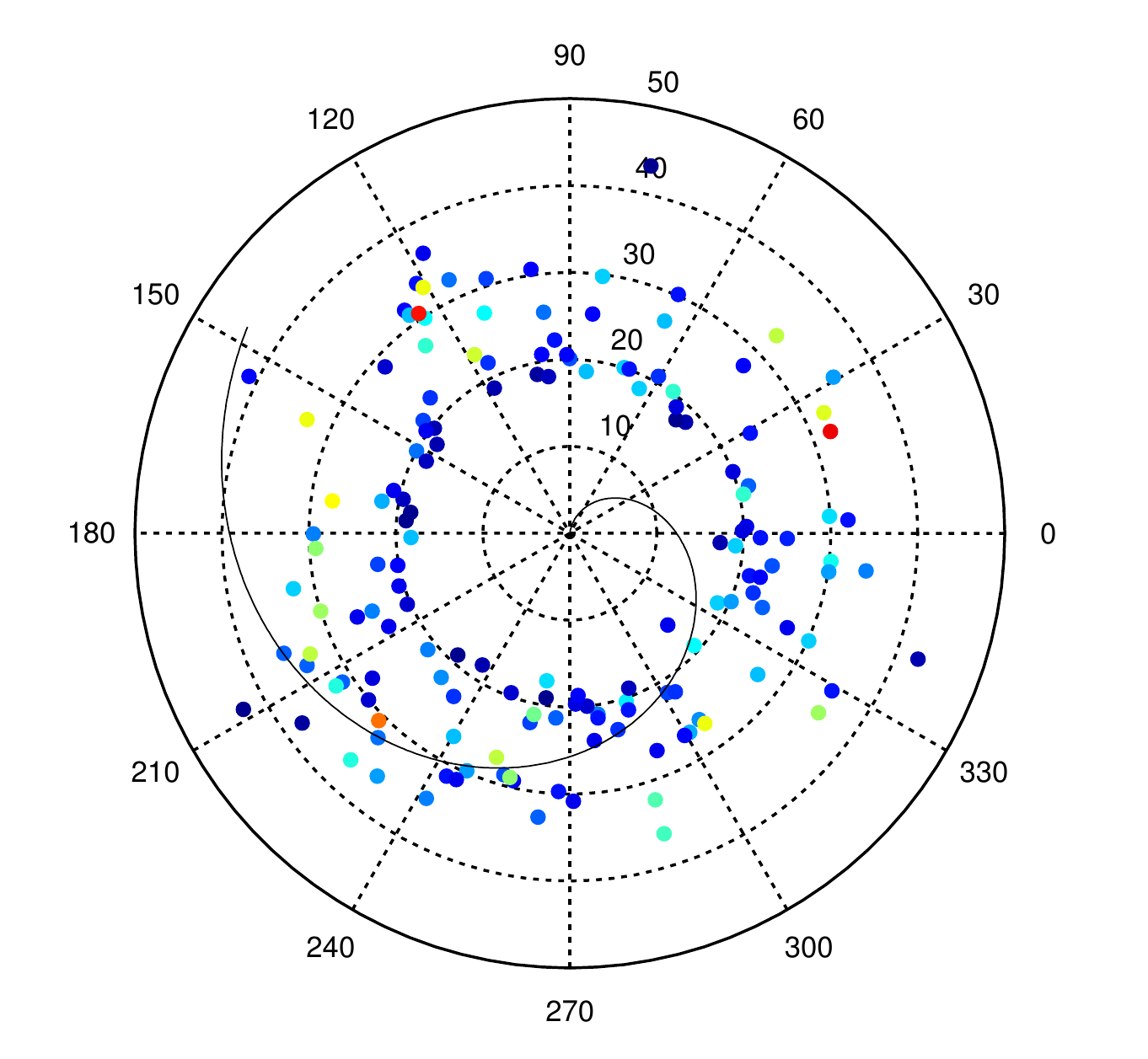}
  (a)
\end{minipage}
\begin{minipage}{.49\textwidth}
  \centering
  \includegraphics[width=20pc]{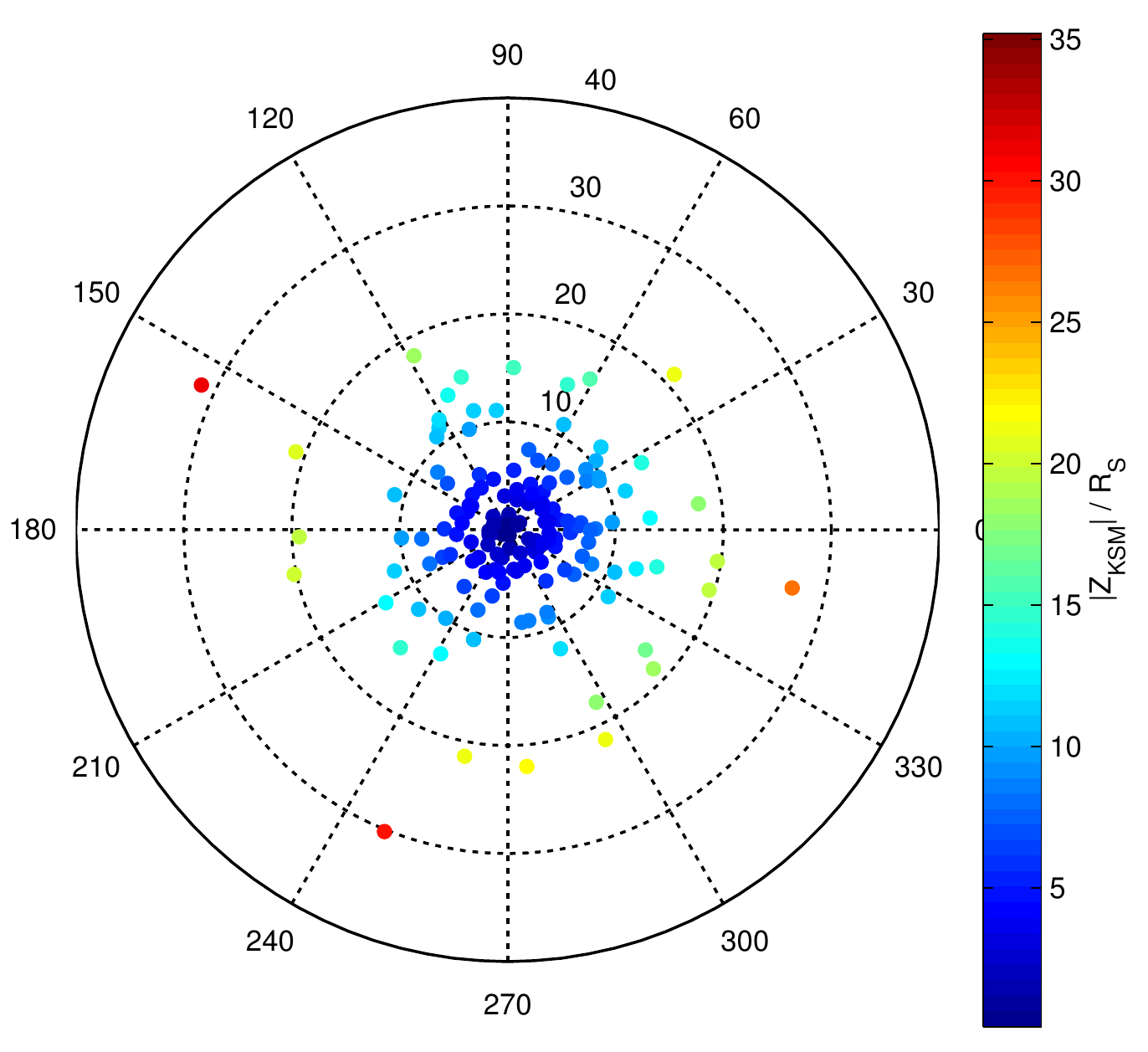}
  (b)
\end{minipage}
\caption{The reduced set of magnetopause crossings have been transformed into the SLS3 longitude system of \protect\citet{Kurth2008}. In (a) the crossings are plotted in this system, the radial distance corresponds to the planet-crossing distance and the crossings are coloured by their normalised $Z_{\textnormal{KSM}}$ coordinates for comparison with other figures. The peak phase front is plotted as a dark line. In (b) the phase difference between each magnetopause crossing and the peak phase front is plotted in order to account for the effects of the bend back of the phase front due to the finite wave speed. The markers surrounding the outermost circle denote the phase difference and the inner markers denote the $Z_{\textnormal{KSM}}$ coordinate of each crossing. There is good coverage of crossings at large $Z_{\textnormal{KSM}}$ (coloured green-red).}
\label{fig:sls3}
\end{figure}

As mentioned previously, the planetary-period magnetic oscillation that has been observed at Saturn is likely to have had an effect on the position of the magnetopause crossings used in this study. For the current study, this oscillation is not taken into account and is treated as noise. \\
However, a small investigation has been undertaken to determine if the apparent flattening of the magnetosphere is caused by this magnetic oscillation. To do this, the SLS3 longitudinal system of \citet{Kurth2008} has been used. The SLS3 longitude of the spacecraft at each crossing has been calculated and plotted in Figure \ref{fig:sls3} (a). The longitude of the peak phase front (defined as 100\,$^\circ$ longitude in the SLS3 system) has also been plotted taking into account bend back effects due to the finite wave speed using parameters determined by \citet{Arridge2011}. Specifically, the distance at which the plasma sheet becomes tilted is taken as 12\,$R_\textnormal{S}$ and the phase delay is taken as 6.7\,$^\circ\,R_\textnormal{S}^{-1}$. The difference in phase between the crossings and the peak phase front have been plotted in Figure \ref{fig:sls3} (b) against the $Z_{\textnormal{KSM}}$ coordinate of the crossing. \\
The key result displayed in Figure \ref{fig:sls3} is that there is no evidence of a pattern between the distribution of the crossings in the SLS3 system and the $Z_{\textnormal{KSM}}$ coordinate. The relevant crossings at large $Z_{\textnormal{KSM}}$ are distributed fairly evenly in SLS3 longitude. If the magnetopause flattening was highly dependant on the magnetic oscillation then it would be expected that the crossings at larger $Z_{\textnormal{KSM}}$ would be clustered together at a similar longitude in the SLS3 system, but this is not the case. \\
In addition to this, if the magnetic oscillation was the dominant influence on the high-latitude boundary location, the normalised boundary locations at large $Z_{\textnormal{KSM}}$ would be scattered evenly around the axisymmetric model boundary. This is not the case as most of these crossings lie inside the axisymmetric surface as can be seen in Figures \ref{fig:x-y_slice} and \ref{fig:x-y_slice_reduced}.

\section{Discussion}
\label{sec:discussion}
We have investigated the structure of the magnetopause of Saturn using in-situ \textit{Cassini} data paying particular attention to the high-latitude regions which haven't previously been studied in detail. Magnetometer and electron plasma spectrometer data has been used to identify magnetopause crossings from a set of highly inclined orbits and estimate the solar wind dynamic pressure at each crossing. \\
This allowed us to normalise the crossings to a fixed pressure so that we could fit models to the data to determine if the magnetosphere of Saturn exhibits polar flattening as has been observed at Jupiter by \citet{Huddleston1998}. Even so, a considerable amount of scatter is present in the data and further measures were taken to reduce this by comparing two different pressure estimates for each crossing and removing those where these estimates deviate by more than a factor of 3. \\
By applying a simple dilation to the axisymmetric magnetopause boundary in the $Z_{KSM}$ direction, a north-south flattening of 19\% within a confidence interval of \hbox{13-22\%} has been found compared to the axisymmetric case. \\
The magnetopause crossings identified in this investigation are limited almost exclusively to the dusk sector of the magnetopause and all of those in dawn sector are located at equatorial latitudes. Future studies should include crossings from the dawn sector such that east-west asymmetries in the structure of the magnetopause can be identified. \\
Seasonal variations are expected due to the hinging effect of the magnetodisc and the north-south asymmetry that this introduces to the magnetic field structure of the magnetosphere. The magnetopause crossings used in this study were located when the planet was approaching the vernal equinox with a dipole tilt angle of \hbox{$\sim$$13^\circ$ - $4^\circ$} . A similar study performed at a different planetary season may reveal what effects the hinging of the magnetodisc has on the structure of the magnetopause. \\
Additional layers of complexity could be added to the model in future studies to improve its fit to the data. The phase of the magnetic oscillation has been briefly touched upon in this study to determine if it could explain the apparent polar flattening that we observe. It was found that the crossings at high $Z_{\textnormal{KSM}}$ where we see a large degree of polar flattening are not at a similar oscillation phase which indicates it is not the cause. However, the oscillation should have some degree of an effect on the location of the magnetopause as found by \citet{Clarke2006a}, and it is likely that if this effect was properly taken into account in future studies, there would be less scatter in the positions of the normalised crossings. \\


%
%
%
%
%
%
%

\begin{acknowledgments}
N.M.P thanks the UK Science and Technology Facilities Council (STFC) for their support (grant code ST/K502406/1) through a PhD studentship.
\end{acknowledgments}

%
%
%
%
%
%
%
%
%

\bibliographystyle{agufull08}
\bibliography{library}



%

%
%
\end{article}
%
%
%
%
%
%

%
%

\begin{table}
\caption{Results of Statistical Tests}
\centering
\begin{tabular}{l c c c c c}
\hline\hline
Wilcoxon Signed Rank Test & $Z_{\textnormal{KSM}}< 5\,R_\textnormal{S}$ & $5<Z_{\textnormal{KSM}}<10\,R_\textnormal{S}$ & $Z_{\textnormal{KSM}}>10\,R_\textnormal{S}$ \\ [0.5ex]
\hline
Unflattened \textit{p}-value & 0.01 (70) & 0.07 (48) & 0.00 (40) \\
Flattened \textit{p}-value & 0.03 (70) & 0.66 (47) & 0.72 (41) \\ [1ex]
\hline
\end{tabular}
\label{tb:stats}
\end{table}

\end{document}